\documentclass[preprint,amsmath,amssymb]{revtex4}

\usepackage{titlesec}
\usepackage{graphicx,epsfig,epstopdf}
\usepackage{bm}
\usepackage{epstopdf}
\usepackage{amssymb,amsmath,amsfonts,latexsym}
\usepackage{dcolumn}
\usepackage{bm}
\usepackage{hyperref}
\usepackage[italic]{hepnames}
\usepackage[utf8]{inputenc}

\usepackage{breqn}             
\makeatletter                  
\let\cat@comma@active\@empty   
\makeatother                   

\begin{document}

\title{$Q\bar Q$ ($Q\in \{b, c\}$) spectroscopy using Cornell potential}%

\author{N. R. Soni}
\email{nrsoni-apphy@msubaroda.ac.in}
\author{B. R. Joshi}
\email{brijaljoshi99@gmail.com}
\author{R. P. Shah}
\email{shahradhika61@gmail.com}
\author{H. R. Chauhan}
\email{hemangichauhan29@gmail.com}
\author{J. N. Pandya}
\email{jnpandya-apphy@msubaroda.ac.in}
\affiliation{Applied Physics Department, Faculty of Technology and Engineering, \\ The Maharaja Sayajirao University of Baroda, Vadodara 390001, Gujarat, India.}

\date{\today}

\begin{abstract}
The mass spectra and decay properties of heavy quarkonia are computed in nonrelativistic quark-antiquark Cornell potential model. We have employed the numerical solution of Schr\"odinger equation to obtain their mass spectra using only four parameters namely quark mass ($m_c$, $m_b$) and confinement strength ($A_{c\bar c}$, $A_{b\bar b}$). The spin hyperfine, spin-orbit and tensor components of the one gluon exchange interaction are computed perturbatively to determine the mass spectra of excited $S$, $P$, $D$ and $F$ states. Digamma, digluon and dilepton decays of these mesons are computed using the model parameters and numerical wave functions. The predicted spectroscopy and decay properties for quarkonia are found to be consistent with available data from experiments, lattice QCD and other theoretical approaches. We also compute mass spectra and life time of the $B_c$ meson without additional parameters. The computed electromagnetic transition widths of heavy quarkonia and $B_c$ mesons are in tune with available experimental data and other theoretical approaches.
\end{abstract}

\pacs{12.38.Bx; 12.39.Pn; 13.20.Gd; 13.40.Hq; 14.40.Pq}
\keywords{Cornell potential, decay properties, electromagnetic transition}
\maketitle
\section{Introduction}
\label{sec:introduction}
Mesonic bound states having both heavy quark and antiquark ($c\bar c$, $b\bar b$ and $c\bar b$) are among the best tools for understanding the quantum chromodynamics. Many experimental groups such as CLEO, LEP, CDF, D0 and NA50 have provided data and \textit{BABAR}, Belle, CLEO-III, ATLAS, CMS and LHCb are producing and expected to produce more precise data in upcoming experiments. Comprehensive reviews on the status of experimental heavy quarkonium physics are found in literature  \cite{Eichten:2007,Godfrey:2008,Barnes:2009,Brambilla:2010,Brambilla:2014,Andronic:2016}.

Within open flavor threshold, the heavy quarkonia have very rich spectroscopy with narrow and experimentally characterized states. The potential between the interacting quarks within the hadrons demands the understanding of underlying physics of strong interactions. In PDG \cite{pdg2016}, large amount of experimental data is available for masses along with different decay modes.
There are many theoretical groups viz. the lattice quantum chromodynamics (LQCD) \cite{Dudek:2007,Meinel:2009,Burch:2009,Liu:2012,McNeile:2012,Daldrop:2011,Kawanai:2013,Kawanai:2011,Burnier:2015,Kalinowski:2015,Burnier:2016}, QCD \cite{Hilger:2014,Voloshin:2007},
QCD sum rules \cite{Cho:2014,Gershtein:1995}, perturbative QCD \cite{Kiyo:2013}, lattice NRQCD \cite{Liu:2016,Dowdall:2013} and
effective field theories \cite{Neubert:1993} that have attempted to explain the production and decays of these states.
Others include phenomenological potential models such as
the relativistic quark model based on quasi-potential approach \cite{Ebert:2011,Ebert:2009,Ebert:2005,Ebert:2002,Ebert:2003lepton,Ebert:2003gamma,Ebert:1999}, where the relativistic quasi-potential including one loop radiative corrections reproduce the mass spectrum of quarkonium states.
The quasi-potential has also been employed along with leading order radiative correction to heavy quark potential \cite{Gupta:1981,Gupta:1982kp,Gupta:1982,Pantaleone:1985}, relativistic potential model \cite{Maung:1993,Radford:2007,Radford:2009} as well as semirelativistic potential model \cite{Gupta:1986}.
In nonrelativistic potential models, there exist several forms of quark antiquark potentials in the literature. The most common among them is the coulomb repulsive plus quark confinement interaction potential. In our previous work  \cite{Vinodkumar:1999,Pandya:2001,Rai:2008,Pandya:2014}, we have employed the confinement scheme based on harmonic approximation along with Lorentz scalar plus vector potential.
The authors of \cite{Devlani:2014,Parmar:2010,Rai:2002,Rai:2005,Rai:2006,Rai:2008prc,Patel:2008} have considered the confinement of power potential $A r^\nu$ with $\nu$ varying from 0.1 to 2.0 and the confinement strength $A$ to vary with potential index $\nu$.
Confinement of the order $r^{2/3}$ have also been attempted \cite{FabreDeLaRipelle:1988}. Linear confinement of quarks has been considered by many groups \cite{Eichten:1974,Eichten:1978,Eichten:1979,Quigg:1979,Eichten:1980,Barnes:2005,Sauli:2011,Leitao:2014,Godfrey:1985,Godfrey:2004,Godfrey:2015,Deng:2016cc,Deng:2016bb} and they have provided good agreement with the experimental data for quarkonium spectroscopy along with decay properties.
The Bethe-Salpeter approach was also employed for the mass spectroscopy of charmonia and bottomonia \cite{Sauli:2011,Leitao:2014,Fischer:2014}.
The quarkonium mass spectrum was also computed in the nonrelativistic quark model \cite{Lakhina:2006}, screened potential model \cite{Deng:2016cc,Deng:2016bb} and constituent quark model \cite{Segovia:2016}.
There are also other non-linear potential models that predict the mass spectra of the heavy quarkonia successfully \cite{Patel:2015,Bonati:2015,Gutsche:2014,Shah:2012,Negash:2015,Bhaghyesh:2011,Li:2009,Li:2009bb,Quigg:1977,Martin:1980,Buchmuller:1980}.

In 90's, the nonrelativistic potential models predicted not only the ground state mass of the tightly bound state of $c$ and $\bar b$ in the range of 6.2--6.3 GeV \cite{Kwong:1990,Eichten:1994} but also predicted to have very rich spectroscopy.
In 1998, CDF collaboration \cite{Abe:1998} reported $B_c$ mesons in $p\bar p$ collisions at $\sqrt{s}$ = 1.8 TeV and was later confirmed by D0 \cite{Abazov:2008} and LHCb \cite{Aaij:2012} collaborations.
The LHCb collaboration has also made the most precise measurement of the life time of $B_c$ mesons \cite{Aaij:2014}.
The first excited state is also reported by ATLAS Collaborations \cite{Aad:2014} in $p\bar p$ collisions with significance of $5.2\sigma$.

It is important to show that any given potential model should be able to compute mass spectra and decay properties of $B_c$ meson using parameters fitted for heavy quarkonia. Attempts in this direction have been made in relativistic quark model based on quasi-potential along with one loop radiative correction \cite{Ebert:2011}, quasistatic and confinement QCD potential with confinement parameters along with quark masses \cite{Gupta:1996} and rainbow-ladder approximation of Dyson-Schwinger and Bethe-Salpeter equations \cite{Fischer:2014}.

The interaction potential for mesonic states is difficult to derive for full range of quark antiquark separation from first principles of QCD. So most forms of QCD inspired potential would result in uncertainties in the computation of spectroscopic properties particularly in the intermediate range. Different potential models may produce similar mass spectra matching with experimental observations but they may not be in mutual agreement when it comes to decay properties like decay constants, leptonic decays or radiative transitions. Moreover, the mesonic states are identified with masses along with certain decay channels, therefore the test for any successful theoretical model is to reproduce the mass spectrum along with decay properties.
Relativistic as well as nonrelativistic potential models have successfully predicted the spectroscopy but they are found to differ in computation of the decay properties  \cite{Quigg:1977,Eichten:1978,Martin:1980,Buchmuller:1980,Gershtein:1995,Rai:2002,Rai:2005,Rai:2006,Rai:2008prc,Parmar:2010}.
In this article, we employ nonrelativistic potential with one gluon exchange (essentially Coulomb like) plus linear confinement (Cornell potential) as this form of the potential is also supported by LQCD \cite{Bali:2000,Bali:2001,Alexandrou:2002}. We solve the Schr\"odinger equation numerically for the potential to get the spectroscopy of the quarkonia. We first compute the mass spectra of charmonia and bottomonia states to determine quark masses and confinement strengths after fitting the spin-averaged ground state masses with experimental data of respective mesons. Using the potential parameters and numerical wave function, we compute the decay properties such as leptonic decay constants,  digamma, dilepton, digluon decay width using the Van-Royen Weiskopf formula. These parameters are then used to compute the mass spectra and life-time of $B_c$ meson. We also compute the electromagnetic ($E1$ and $M1$) transition widths of heavy quarkonia and $B_c$ mesons.

\section{Methodology}
\label{sec:methodlogy}
Bound state of two body system within relativistic quantum field is described in Bethe-Salpeter formalism. However, the Bethe-Salpeter equation is solved only in the ladder approximations. Also, Bethe-Salpeter approach in harmonic confinement is successful in low flavor sectors \cite{Isgur:1978,VijayaKumar:1997}. Therefore the alternative treatment for the heavy bound state is nonrelativistic. Significantly low momenta of quark and antiquark compared to mass of quark-antiquark system $m_{Q,\bar{Q}} \gg \Lambda_{QCD} \sim	|\vec{p}|$ also constitutes the basis of the nonrelativistic treatment for the heavy quarkonium spectroscopy. Here, for the study of heavy bound state of mesons such as $c\bar{c}$, $c\bar{b}$ and $b\bar{b}$, the nonrelativistic Hamiltonian is given by
\begin{equation}\label{eq:hamiltonian}
H = M + \frac{p^2}{2 M_{cm}} + V_{{\text{Cornell}}}(r) + V_{SD} (r)
\end{equation}
where
\begin{equation}
M = m_Q + m_{\bar{Q}}  \ \ \ \  \text{and} \ \  \ \ M_{cm} = \frac{m_Q m_{\bar{Q}}} {m_{Q} + m_{\bar{Q}}}
\end{equation}
where $m_Q$ and $m_{\bar{Q}}$ are the masses of quark and antiquark respectively, $\vec{p}$ is the relative momentum of the each quark and $V_{\text{Cornell}} (r)$ is the quark-antiquark potential of the type coulomb plus linear confinement (Cornell potential) given by
\begin{eqnarray}\label{eq_cornell}
    V_{\text{Cornell}}(r) = - \frac{4}{3}\frac{\alpha_s}{r} + A r .
\end{eqnarray}
Here, $1/r$ term is analogous to the Coulomb type interaction corresponding to the potential induced between quark and antiquark through one gluon exchange that  dominates at small distances. The second term is the confinement part of the potential with the confinement strength $A$ as the model parameter. The confinement term becomes dominant at the large distances. $\alpha_s$ is a strong running coupling constant and can be computed as
\begin{eqnarray}\label{eq:running_coupling}
    \alpha_s (\mu^2) = \frac{4 \pi}{(11-\frac{2}{3} n_f) \ln (\mu^2/\Lambda^2)}
\end{eqnarray}
where $n_f$ is the number of flavors, $\mu$ is renormalization scale related to the constituent quark masses as $\mu = 2 m_Q m_{\bar Q}/(m_Q + m_{\bar Q})$ and $\Lambda$ is a QCD scale which is taken as 0.15 GeV by fixing $\alpha_s$ = 0.1185 \cite{pdg2016} at the $Z$-boson mass.

The confinement strengths with respective quark masses are fine tuned to reproduce the experimental spin averaged ground state masses of both $c\bar{c}$ and $b\bar{b}$ mesons and they are given in Table \ref{tab:parameters}. We compute the masses of radially and orbitally excited states without any additional parameters. Similar work has been done by \cite{Patel:2008,Rai:2008prc,Parmar:2010} and they have considered different values of confinement strengths for different potential indices. The Cornell potential has been shown to be  independently successful in computing the spectroscopy of $\psi$ and $\Upsilon$ families. In this article, we compute the mass spectra of the $\psi$ and $\Upsilon$ families along with $B_c$ meson with minimum number of parameters.

Using the parameters defined in Table \ref{tab:parameters}, we compute the spin averaged masses of quarkonia. In order to compute masses of different $n^m L_J$ states according to different $J^{PC}$ values, we use the spin dependent part of one gluon exchange potential (OGEP) $V_{SD} (r)$ perturbatively.  The OGEP includes spin-spin, spin-orbit and tensor terms given by  \cite{Gershtein:1995,Barnes:2005,Lakhina:2006,Voloshin:2007}
\begin{dmath}\label{eq:vsd}
    V_{SD} (r) = V_{SS} (r) \left[S(S+1) - \frac{3}{2}\right] + V_{LS}(r) (\vec L\cdot\vec S) + V_T(r) \left[S(S+1)-3(S\cdot \hat r) (S\cdot \hat r)\right]
\end{dmath}
\begin{table}[h]
\caption{Parameters for quarkonium spectroscopy}\label{tab:parameters}
\begin{tabular}{ccccccc}
$m_c$ & $m_c$ & $A_{cc}$ & $A_{bb}$\\
\hline
1.317 GeV & 4.584 GeV & 0.18 GeV$^2$ & 0.25 GeV$^2$\\
\end{tabular}
\end{table}

The spin-spin interaction term gives the hyper-fine splitting while spin-orbit and tensor terms gives the fine structure of the quarkonium states. The coefficients of spin dependent terms of the Eq. (\ref{eq:vsd}) can be written as \cite{Voloshin:2007}
\begin{table*}[htbp]
\caption{Mass spectrum of $S$ and $P$-wave charmonia (in GeV)}\label{tab:cc_sp_mass}
\begin{tabular*}{\textwidth}{@{\extracolsep{\fill}}ccccccccccccc@{}}
\hline
State    &Present &\cite{Ebert:2011}&\cite{Deng:2016cc}&\cite{Fischer:2014} & \cite{Li:2009} & \cite{Radford:2007}& \cite{Shah:2012} & \cite{Barnes:2005} & \cite{Lakhina:2006}&\cite{Patel:2015} &LQCD \cite{Kalinowski:2015} & PDG \cite{pdg2016}\\
\hline
$1^1S_0$ & 2.989	 & 2.981 & 2.984	& 2.925 & 2.979 & 2.980 	& 2.980	& 2.982 & 3.088	& 2.979 & 2.884 & 2.984\\
$1^3S_1$ & 3.094  & 3.096 & 3.097	& 3.113 & 3.097 & 3.097 	& 3.097	& 3.090 & 3.168	& 3.096 & 3.056 & 3.097\\
\hline
$2^1S_0$ & 3.602  & 3.635 & 3.637	& 3.684 & 3.623 & 3.597 	& 3.633	& 3.630 & 3.669	& 3.600 & 3.535 & 3.639\\
$2^3S_1$ & 3.681  & 3.685 & 3.679	& 3.676 & 3.673 & 3.685 	& 3.690	& 3.672 & 3.707	& 3.680 & 3.662 & 3.686\\
\hline
$3^1S_0$ & 4.058  & 3.989 & 4.004	& --		& 3.991 & 4.014 	& 3.992	& 4.043 & 4.067	& 4.011  & --		& --\\
$3^3S_1$ & 4.129  & 4.039 & 4.030	& 3.803 & 4.022 & 4.095	& 4.030	& 4.072 & 4.094	& 4.077 & --		& 4.039\\
\hline
$4^1S_0$ & 4.448  & 4.401 & 4.264	& --	 	& 4.250 & 4.433		& 4.244	& 4.384 & 4.398	& 4.397 & -- 		& --\\
$4^3S_1$ & 4.514	 & 4.427 & 4.281	& -- 		& 4.273 & 4.477		& 4.273	& 4.406 & 4.420	&4.454 & -- 		& 4.421\\
\hline
$5^1S_0$ & 4.799  & 4.811 & 4.459	& --    	& 4.446 & --			& 4.440	& --		&  --		& -- 		& --		& -- \\
$5^3S_1$ & 4.863  & 4.837 & 4.472	& --    	& 4.463 & --			& 4.464	& --		&  --		& -- 		&-- 		& --\\
\hline
$6^1S_0$ & 5.124  & 5.155 & --			& --    	& 4.595 & --			& 4.601	& --		&  --		& -- 		& -- 		& --\\
$6^3S_1$ & 5.185  & 5.167 & --			& --    	& 4.608 & -- 			& 4.621	& --		& --		& -- 		& --		& --\\
\hline
$1^3P_0$ & 3.428  & 3.413 & 3.415	& 3.323 & 3.433 & 3.416	& 3.392	& 3.424 & 3.448	& 3.488 & 3.412 & 3.415\\
$1^3P_1$ & 3.468  & 3.511 & 3.521	& 3.489 & 3.510 & 3.508	& 3.491	& 3.505	& 3.520	& 3.514 & 3.480 & 3.511\\
$1^1P_1$ & 3.470  & 3.525 & 3.526	& 3.433 & 3.519 & 3.527	& 3.524	& 3.516	& 3.536	& 3.539 & 3.494 & 3.525\\
$1^3P_2$ & 3.480  & 3.555 & 3.553	& 3.550 & 3.556 & 3.558	& 3.570	& 3.556 & 3.564	& 3.565 & 3.536 & 3.556\\
\hline
$2^3P_0$ & 3.897  & 3.870 & 3.848	& 3.833 & 3.842 & 3.844	& 3.845	& 3.852 & 3.870	& 3.947 & --		& 3.918\\
$2^3P_1$ & 3.938	 & 3.906 & 3.914	& 3.672 & 3.901 & 3.940	& 3.902	& 3.925 & 3.934	& 3.972 & --		& --\\
$2^1P_1$ & 3.943  & 3.926 & 3.916	& 3.747 & 3.908 & 3.960	& 3.922	& 3.934 & 3.950	& 3.996 & --		& --\\
$2^3P_2$ & 3.955 & 3.949 & 3.937	& --		& 3.937 & 3.994		& 3.949	& 3.972 & 3.976	& 4.021 & 4.066	& 3.927\\
\hline
$3^3P_0$ & 4.296 & 4.301 & 4.146	& --    	& 4.131	 & --			& 4.192	& 4.202 & 4.214	& --		 & --		& --\\
$3^3P_1$ & 4.338 & 4.319 & 4.192	& 3.912	& 4.178	 & --			& 4.178	& 4.271 & 4.275	& --		 & --		& --\\
$3^1P_1$ & 4.344 & 4.337 & 4.193	& --    	& 4.184	 & --			& 4.137	& 4.279 & 4.291	& --		& --		& --\\
$3^3P_2$ & 4.358 & 4.354 & 4.211		& --    	& 4.208	 & --			& 4.212	& 4.317 & 4.316	& --		 & --		& --\\
\hline
$4^3P_0$ & 4.653 & 4.698 & --			& --    	& --		 & --			& --		& --		& --		& -- 		& --		& --\\
$4^3P_1$ & 4.696 & 4.728 & --			& --    	& --		 & --			& --		& --		& --		& -- 		& --		& --\\
$4^1P_1$ & 4.704 & 4.744 & --			& --    	& --		 & --			& --		& --		& --		& -- 		& --		& --\\
$4^3P_2$ & 4.718 & 4.763 & --			& --    	& --		 & --			& --		& --		& --		& -- 		& --		& --\\
\hline
$5^3P_0$ & 4.983 & --       	& --		& --    	& -- 		 & --			& --		& --		& --		& -- 		& --		& --\\
$5^3P_1$ & 5.026 & --       	& --		& --    	& -- 		 & --			& --		& --		& --		& -- 		& --		& --\\
$5^1P_1$ & 5.034 & --       	& --		& --    	& -- 		 & --			& --		& --		& --		& -- 		& --		& --\\
$5^3P_2$ & 5.049 & --       	& --		& --    	& -- 		 & --			& --		& --		& --		& -- 		& --		& --
\\ \hline \end{tabular*}
\end{table*}
\begin{eqnarray}
  V_{SS} (r) = \frac{1}{3 m_Q m_{\bar Q}} \nabla^2 V_V(r) = \frac{16 \pi \alpha_s}{9 m_Q m_{\bar Q}} \delta^3(\vec{r})
\end{eqnarray}
\begin{eqnarray}
V_{LS} (r) = \frac{1}{2 m_Q m_{\bar Q} r} \left(3 \frac{dV_V(r)}{dr} - \frac{dV_S(r)}{dr}\right)
\end{eqnarray}
\begin{eqnarray}
V_T(r) = \frac{1}{6 m_Q m_{\bar Q}} \left(3 \frac{dV^2_V(r)}{dr^2} -\frac{1}{r} \frac{dV_V(r)}{dr}\right)
\end{eqnarray}
Where $V_V(r)$ and $V_S(r)$ correspond to the vector and scalar part of the Cornell potential in Eq. (\ref{eq_cornell}) respectively. Using all the parameters defined above, the Schr\"odinger equation is numerically solved using {\it Mathematica} notebook utilizing the Runge-Kutta method \cite{Lucha:1998}. It is generally believed that the charmonia need to be treated relativistically due to their lighter masses, but we note here that the computed wave functions of charmonia using relativistic as well as nonrelativistic approaches don't show significant difference \cite{Ebert:1999}. So we choose to compute the charmonium mass spectra nonrelativistically in present study. The computed mass spectra of heavy quarkonia and $B_c$ mesons are listed in Tables \ref{tab:cc_sp_mass}--\ref{tab:bc_df_mass}.
\begin{table*}[htbp]
\caption{Mass spectrum of $D$ and $F$-wave charmonia (in GeV)}\label{tab:cc_df_mass}
\begin{tabular*}{\textwidth}{@{\extracolsep{\fill}}ccccccccccc@{}}
\hline
State    &Present & \cite{Ebert:2011}&\cite{Deng:2016cc}&\cite{Fischer:2014}  &\cite{Li:2009}&  \cite{Radford:2007} & \cite{Shah:2012} & \cite{Barnes:2005} &\cite{Lakhina:2006} & \cite{Patel:2015}\\
\hline
$1^3D_3$ & 3.755     & 3.813 & 3.808 & 3.869 & 3.799	& 3.831	& 3.844	& 3.806 & 3.809 & 3.798\\
$1^1D_2$ & 3.765     & 3.807 & 3.805 & 3.739 & 3.796	& 3.824	& 3.802	& 3.799 & 3.803 & 3.796\\
$1^3D_2$ & 3.772     & 3.795 & 3.807 & 3.550 & 3.798	& 3.824	& 3.788	& 3.800 & 3.804	& 3.794\\
$1^3D_1$ & 3.775     & 3.783 & 3.792 & -- 		 & 3.787	& 3.804	& 3.729	& 3.785 & 3.789 & 3.792\\
\hline
$2^3D_3$ & 4.176     & 4.220 & 4.112	 & 3.806 & 4.103	& 4.202	& 4.132	& 4.167 & 4.167 & 4.425\\
$2^1D_2$ & 4.182     & 4.196 & 4.108 & --		 & 4.099	& 4.191	& 4.105	& 4.158 & 4.158 & 4.224\\
$2^3D_2$ & 4.188     & 4.190 & 4.109 & --		 & 4.100	& 4.189	& 4.095	& 4.158 & 4.159 & 4.223\\
$2^3D_1$ & 4.188     & 4.105 & 4.095 & -- 		 & 4.089	& 4.164	& 4.057	& 4.142 & 4.143 & 4.222\\
\hline
$3^3D_3$ & 4.549     & 4.574 & 4.340 & --    	 & 4.331	& --		& 4.351	& --		& --		 & -- \\
$3^1D_2$ & 4.553     & 3.549 & 4.336 & --    	 & 4.326	& --		& 4.330	& --		& --		 & -- \\
$3^3D_2$ & 4.557     & 4.544 & 4.337 & --    	 & 4.327	& --		& 4.322	& --		& --		 & -- \\
$3^3D_1$ & 4.555     & 4.507 & 4.324 & --    	 & 4.317	& --		& 4.293	& --		& --		 & -- \\
\hline
$4^3D_3$ & 4.890     & 4.920 & --		 & --    	 & --			& --		& 4.526	& --		& --		 & -- \\
$4^1D_2$ & 4.892     & 4.898 & --		 & --    	 & --			& --		& 4.509	& --		& --		 & -- \\
$4^3D_2$ & 4.896     & 4.896 & --		 & --    	 & --			& --		& 4.504	& --		& --		 & -- \\
$4^3D_1$ & 4.891     & 4.857 & --		 & --    	 & --			& --		& 4.480	& --		& --		 & -- \\
\hline
$1^3F_2$ & 3.990     & 4.041 & --		 & --    	 & --			& 4.068	& --		& 4.029 & --		 & --  \\
$1^3F_3$ & 4.012     & 4.068 & --		 & 3.999 & --			& 4.070	& --		& 4.029 & --		 & -- \\
$1^1F_3$ & 4.017     & 4.071 & --		 & 4.037 & --			& 4.066	& --		& 4.026 & --		 & -- \\
$1^3F_4$ & 4.036     & 4.093 & --		 & --    	 & --			& 4.062	& --		& 4.021 & --		 & -- \\
\hline
$2^3F_2$ & 4.378     & 4.361 & --		 & --    	& --			& --		& --		& 4.351 & --		 & -- \\
$2^3F_3$ & 4.396     & 4.400 & --		 & --    	& --			& --		& --		& 3.352 & --		 & -- \\
$2^1F_3$ & 4.400     & 4.406 & --		 & --    	& --			& --		& --		& 4.350 & --		 & -- \\
$2^3F_4$ & 4.415     & 4.434 & --		 & --    	& --			& --		& --		& 4.348 & --		 & -- \\
\hline
$3^3F_2$ & 4.730     & --    	& --		 & --    	& --			& --		& --		& --		& --		 & --  \\
$3^3F_3$ & 4.746     & --    	& --		 & --    	& --			& --		& --		& --		& --		 & -- \\
$3^1F_3$ & 4.749     & --    	& --		 & --    	& --			& --		& --		& --		& --		 & -- \\
$3^3F_4$ & 4.761     & --    	& --		 & --    	& --			& --		& --		& --		& --		 & --
\\ \hline \end{tabular*}
\end{table*}
\begin{table*}[htbp]
\caption{Mass spectrum of $S$ and $P$-wave bottomonia (in GeV)}\label{tab:bb_sp_mass}
\begin{tabular*}{\textwidth}{@{\extracolsep{\fill}}ccccccccccc@{}}
\hline
State & Present  & \cite{Godfrey:2015}&\cite{Ebert:2011}	&\cite{Deng:2016bb} & \cite{Fischer:2014} &\cite{Li:2009bb}&  \cite{Radford:2009} &  \cite{Shah:2012} & \cite{Segovia:2016} & PDG \cite{pdg2016}\\
\hline
$1^1S_0$& 9.428		& 9.402 	& 9.398  	& 9.390  	& 9.414 	& 9.389 	& 9.393		& 9.392 	& 9.455 	& 9.398\\
$1^3S_1$& 9.463  	& 9.465 	& 9.460  	& 9.460  	& 9.490		& 9.460 	& 9.460		& 9.460		& 9.502 	& 9.460\\
\hline
$2^1S_0$& 9.955  	& 9.976		& 9.990   	& 9.990  	& 9.987		& 9.987 	& 9.987		& 9.991		& 9.990 	& 9999\\
$2^3S_1$& 9.979  	& 10.003	&10.023  	& 10.015	& 10.089	& 10.016	& 10.023	& 10.024	& 10.015 	& 10.023\\
\hline
$3^1S_0$& 10.338	& 10.336	& 10.329 	& 10.326 	& --			& 10.330	& 10.345	& 10.323	& 10.330 	& --\\
$3^3S_1$& 10.359 	& 10.354	& 10.355 	& 10.343	& 10.327	& 10.351 	& 10.364	& 10.346	& 10.349 	& 10.355\\
\hline
$4^1S_0$& 10.663 	& 10.523	& 10.573 	& 10.584   & --			& 10.595 	& 10.623	& 10.558	& -- 			& --\\
$4^3S_1$& 10.683 	& 10.635	& 10.586 	& 10.597   & --			& 10.611 	& 10.643	& 10.575	& 10.607 	& 10.579\\
\hline
$5^1S_0$& 10.956 	& 10.869	& 10.851 	& 10.800   & --			& 10.817 	& --			& 10.741	& -- & --	\\
$5^3S_1$& 10.975 	& 10.878	& 10.869 	& 10.811   	& --			& 10.831	& --			& 10.755	& 10.818 	& 10.876\\
\hline
$6^1S_0$& 11.226 	& 11.097	& 11.061 	& 10.997   & --			& 11.011	& --			& 10.892	& -- & --	\\
$6^3S_1$& 11.243 	& 11.102	& 11.088 	& 10.988   & --			& 11.023	& --			& 10.904	& 10.995 & 11.019\\
\hline
$1^3P_0$& 9.806  	& 9.847		& 9.859  	& 9.864  	& 9.815		& 9.865		& 9.861 	& 9.862		& 9.855 	& 9.859\\
$1^3P_1$& 9.819  	& 9.876		& 9.892  	& 9.903  	& 9.842		& 9.897		& 9.891		& 9.888		& 9.874 	& 9.893\\
$1^1P_1$& 9.821  	& 9.882		& 9.900  	& 9.909    	& 9.806		& 9.903		& 9.900		& 9.896		& 9.879 	& 9.899\\
$1^3P_2$& 9.825		& 9.897		& 9.912  	& 9.921 	& 9.906		& 9.918		& 9.912		& 9.908		& 9.886 	& 9.912\\
\hline
$2^3P_0$& 10.205 	& 10.226	& 10.233 	& 10.220	& 10.254	& 10.226	& 10.230	& 10.241	& 10.221 	& 10.232\\
$2^3P_1$& 10.217 	& 10.246	& 10.255 	& 10.249 	& 10.120	& 10.251	& 10.255	& 10.256	& 10.236 	& 10.255\\
$2^1P_1$& 10.220 	& 10.250	& 10.260 	& 10.254 	& 10.154	& 10.256	& 10.262	& 10.261	& 10.240 	& 10.260\\
$2^3P_2$& 10.224 	& 10.261	& 10.268 	& 10.264 	& --			& 10.269	& 10.271	& 10.268	& 10.246 	& 10.269\\
\hline
$3^3P_0$& 10.540 	& 10.552	& 10.521 	& 10.490   & --			& 10.502	& --			& 10.511	& 10.500	& --\\
$3^3P_1$& 10.553 	& 10.538	& 10.541 	& 10.515   & 10.303	& 10.524	& --			& 10.507	& 10.513	& --\\
$3^1P_1$& 10.556 	& 10.541	& 10.544 	& 10.519  	& --			& 10.529	& --			& 10.497	& 10.516	& --\\
$3^3P_2$& 10.560 	& 10.550	& 10.550 	& 10.528  	&  --			& 10.540	& --			& 10.516	& 10.521	& --\\
\hline
$4^3P_0$& 10.840 	& 10.775	& 10.781 	& --     		& --			& 10.732	& --			& --			& --			& --\\
$4^3P_1$& 10.853 	& 10.788	& 10.802 	& --     		& --			& 10.753	& --			& --			& --			& --\\
$4^1P_1$& 10.855 	& 10.790	& 10.804 	& --     		& --			& 10.757	& --			& --			& --			& --\\
$4^3P_2$& 10.860 	& 10.798	& 10.812 	& --     		& --			& 10.767	& --			& --			& --			& --\\
\hline
$5^3P_0$& 11.115 	& 11.004	& --		 	& --     		& --			& 10.933	& --			& --			& --			& --\\
$5^3P_1$& 11.127 	& 11.014	& --		 	& --     		& --			& 10.951	& --			& --			& -- 			& --\\
$5^1P_1$& 11.130 	& 11.016	& --		 	& --     		& --			& 10.955	& --			& --			& --			& --\\
$5^3P_2$& 11.135 	& 11.022	& --		 	& --     		& --			& 10.965	& --			& --			& --			& --		
\\ \hline \end{tabular*}
\end{table*}
\begin{table*}[htbp]
\caption{Mass spectrum of $D$ and $F$-wave bottomonia (in GeV)}\label{tab:bb_df_mass}
\begin{tabular*}{\textwidth}{@{\extracolsep{\fill}}ccccccccccc@{}}
\hline
State & Present &\cite{Godfrey:2015}&\cite{Ebert:2011}&\cite{Deng:2016bb}&  \cite{Fischer:2014} &\cite{Li:2009bb}&  \cite{Radford:2009} &  \cite{Shah:2012} & \cite{Segovia:2016}  & PDG \cite{pdg2016}\\
\hline
$1^3D_3$& 10.073	& 10.115	& 10.166 	& 10.157 	& 10.232	& 10.156	& 10.163	& 10.177	& 10.127	& --\\
$1^1D_2$& 10.074    & 10.148	& 10.163 	& 10.153 	& 10.194	& 10.152	& 10.158	& 10.166	& 10.123	& --\\
$1^3D_2$& 10.075  	& 10.147	& 10.161 	& 10.153 	& 10.145	& 10.151	& 10.157	& 10.162	& 10.122 & 10.163\\
$1^3D_1$& 10.074    & 10.138	& 10.154 	& 10.146 	& --			& 10.145	& 10.149	& 10.147	& 10.117	& --\\
\hline
$2^3D_3$& 10.423    & 10.455	& 10.449 	& 10.436 	& --			& 10.442	& 10.456	& 10.447	& 10.422	& --\\
$2^1D_2$& 10.424    & 10.450	& 10.445 	& 10.432 	& --			& 10.439	& 10.452	& 10.440	& 10.419	& --\\
$2^3D_2$& 10.424    & 10.449	& 10.443 	& 10.432 	& --			& 10.438	& 10.450	& 10.437	& 10.418	& --\\
$2^3D_1$& 10.423    & 10.441	& 10.435 	& 10.425 	& --			& 10.432	& 10.443	& 10.428	& 10.414	& --\\
\hline
$3^3D_3$& 10.733    & 10.711	& 10.717 	& --    		& --			& 10.680	& --			& 10.652	& --			& --\\
$3^1D_2$& 10.733    & 10.706	& 10.713 	& --    		& --			& 10.677	& --			& 10.646	& --			& --\\
$3^3D_2$& 10.733    & 10.705	& 10.711 	& --    		& --			& 10.676	& --			& 10.645	& --			& --\\
$3^3D_1$& 10.731    & 10.698	& 10.704 	& --    		& --			& 10.670	& --			& 10.637	& --			& --\\
\hline
$4^3D_3$& 11.015	& 10.939	& 10.963 	& --			& --			& 10.886	& --			& 10.817	& --			& --\\
$4^1D_2$& 11.015 	& 10.935	& 10.959 	& --			& --			& 10.883	& --			& 10.813	& --			& --\\
$4^3D_2$& 11.016 	& 10.934	& 10.957 	& --			& --			& 10.882	& --			& 10.811	& --			& --\\
$4^3D_1$& 11.013 	& 10.928	& 10.949 	& --			& --			& 10.877	& --			& 10.805	& --			& --\\
\hline
$1^3F_2$& 10.283    & 10.350	& 10.343 	& 10.338   & --			& --			& 10.353	& --					& 10.315	& --\\
$1^3F_3$& 10.287    & 10.355	& 10.346 	& 10.340   & 10.302	& --			& 10.356	& --					& 10.321	& --\\
$1^1F_3$& 10.288    & 10.355	& 10.347 	& 10.339   & 10.319	& --			& 10.356	& --					& 10.322	& --\\
$1^3F_4$& 10.291    & 10.358	& 10.349 	& 10.340   & --			& --			& 10.357	& --					& --			& --\\
\hline
$2^3F_2$& 10.604    & 10.615	& 10.610 	& --    		& --			& --			& 10.610	& --					& --			& --\\
$2^3F_3$& 10.607    & 10.619	& 10.614 	& --    		& --			& --			& 10.613	& --					& --			& --\\
$2^1F_3$& 10.607    & 10.619	& 10.647 	& --    		& --			& --			& 10.613	& --					& --			& --\\
$2^3F_4$& 10.609    & 10.622	& 10.617 	& --    		& --			& --			& 10.615	& --					& --			& --\\
\hline
$3^3F_2$& 10.894    & 10.850	& -- 			& --     		& --			& --			& --			& --					& --			& --\\
$3^3F_3$& 10.896    & 10.853	& --    		& --     		& --			& --			& --			& --					& --			& --\\
$3^1F_3$& 10.897    & 10.853	& --    		& --     		& --			& --			& --			& --					& --			& --\\
$3^3F_4$& 10.898    & 10.856	& --    		& --     		& --			& --			& --			& --					&  --			& --		
\\ \hline \end{tabular*}
\end{table*}
\begin{table}[htbp]
\caption{Mass spectrum of $S$ and $P$-wave $B_c$ meson (in GeV)}\label{tab:bc_sp_mass}
\begin{tabular*}{\columnwidth}{@{\extracolsep{\fill}}ccccccc@{}}
\hline
State    & Present &\cite{Devlani:2014}& \cite{Ebert:2011}& \cite{Godfrey:2004} & \cite{Monteiro:2016} & PDG \cite{pdg2016} \\
\hline
$1^1S_0$ & 6.272 & 6.278 & 6.272 & 6.271 & 6.275 & 6.275\\
$1^3S_1$ & 6.321 & 6.331 & 6.333 & 6.338 & 6.314 & --\\
\hline
$2^1S_0$ & 6.864 & 6.863 & 6.842 & 6.855 & 6.838 & 6.842 \\
$2^3S_1$ & 6.900 & 6.873 & 6.882 & 6.887 & 6.850 & --\\
\hline
$3^1S_0$ & 7.306 & 7.244 & 7.226 & 7.250 & --		 & --\\
$3^3S_1$ & 7.338 & 7.249 & 7.258 & 7.272 & --		 &  --\\
\hline
$4^1S_0$ & 7.684 & 7.564 & 7.585 & --		 & --		 & --\\
$4^3S_1$ & 7.714 & 7.568 & 7.609 & --		 & --		 & --\\
\hline
$5^1S_0$ & 8.025 & 7.852 & 7.928 & --		 & --		 & --\\
$5^3S_1$ & 8.054 & 7.855 & 7.947 & --		 & --		 & --\\
\hline
$6^1S_0$ & 8.340 & 8.120 & --		 & --		 & --    & -- \\
$6^3S_1$ & 8.368 & 8.122 & --		 & --		 & --    & --\\
\hline
$1^3P_0$ & 6.686 & 6.748 & 6.699 & 6.706 & 6.672 & --\\
$1^3P_1$ & 6.705 & 6.767 & 6.750 & 6.741 & 6.766 & --\\
$1^1P_1$ & 6.706 & 6.769 & 6.743 & 6.750 & 6.828 & --\\
$1^3P_2$ & 6.712 & 6.775 & 6.761 & 6.768 & 6.776 & --\\
\hline
$2^3P_0$ & 7.146 & 7.139 & 7.094 & 7.122 & 6.914 & --\\
$2^3P_1$ & 7.165 & 7.155 & 7.134 & 7.145 & 7.259 & --\\
$2^1P_1$ & 7.168 & 7.156 & 7.094 & 7.150 & 7.322 & --\\
$2^3P_2$ & 7.173 & 7.162 & 7.157 & 7.164 & 7.232 & --\\
\hline
$3^3P_0$ & 7.536 & 7.463 & 7.474 & --    	 & --		 & --\\
$3^3P_1$ & 7.555 & 7.479 & 7.510 & --    	 & -- 		 & --\\
$3^1P_1$ & 7.559 & 7.479 & 7.500 & --    	 & --		 & --\\
$3^3P_2$ & 7.565 & 7.485 & 7.524 & --    	 & --		 & --\\
\hline
$4^3P_0$ & 7.885 & -- 		 & 7.817 & --		 & --		& --\\
$4^3P_1$ & 7.905 & -- 		 & 7.853 & --		 & --		& --\\
$4^1P_1$ & 7.908 & -- 		 & 7.844 & --		 & --		& -- \\
$4^3P_2$ & 7.915 & -- 		 & 7.867 & --		 & --		& --\\
\hline
$5^3P_0$ & 8.207 & --    & --    		 & --    	& --\\
$5^3P_1$ & 8.226 & --    & --    		 & --    	& --\\
$5^1P_1$ & 8.230 & --    & --    		 & --    	& --\\
$5^3P_2$ & 8.237 & --    & --    		 & --    	& --
\\ \hline \end{tabular*}
\end{table}
\begin{table}[htbp]
\caption{Mass spectrum of $D$ and $F$-wave $B_c$ meson (in GeV)}\label{tab:bc_df_mass}
\begin{tabular*}{\columnwidth}{@{\extracolsep{\fill}}cccccc@{}}
\hline
State    & Present &\cite{Devlani:2014}& \cite{Ebert:2011}& \cite{Godfrey:2004} & \cite{Monteiro:2016} \\
\hline
$1^3D_3$ & 6.990 & 7.026 & 7.029 & 7.045	& 6.980 \\
$1^1D_2$ & 6.994 & 7.035 & 7.026 & 7.041 	& 7.009 \\
$1^3D_2$ & 6.997 & 7.025 & 7.025 & 7.036 	& 7.154 \\
$1^3D_1$ & 6.998 & 7.030 & 7.021 & 7.028	& 7.078 \\
\hline
$2^3D_3$ & 7.399 & 7.363 & 7.405 & --    		& -- \\
$2^1D_2$ & 7.401 & 7.370 & 7.400 & --    		& -- \\
$2^3D_2$ & 7.403 & 7.361 & 7,399 & --    		& -- \\
$2^3D_1$ & 7.403 & 7.365 & 7.392 & --    		& -- \\
\hline
$3^3D_3$ & 7.761 & --    	& 7.750 &   --    		& --    \\
$3^1D_2$ & 7.762 & --    	& 7.743 &   --    		& --    \\
$3^3D_2$ & 7.764 & --    	& 7.741 &   --    		& --    \\
$3^3D_1$ & 7.762 & --    	& 7.732 &   --    		& --    \\
\hline
$4^3D_3$ & 8.092	& --		& --		& --			&--	\\
$4^1D_2$ & 8.093 & --		& --		& --			&--	\\
$4^3D_2$ & 8.094 & --		& --		& --			&--	\\
$4^3D_1$ & 8.091 & --		& --		& --			&--	\\
\hline
$1^3F_2$ & 7.234 & --    	& 7.273 & 7.269    	& --    \\
$1^3F_3$ & 7.242 & --    	& 7.269 & 7.276    	& --    \\
$1^1F_3$ & 7.241 & --    	& 7.268 & 7.266    	& --    \\
$1^3F_4$ & 7.244 & --    	& 7.277 & 7.271	 	& --    \\
\hline
$2^3F_2$ & 7.607 & --    	& 7.618 & --    		& --    \\
$2^3F_3$ & 7.615 & --    	& 7.616 & --    		& --    \\
$2^1F_3$ & 7.614 & --    	& 7.615	& --    		& --    \\
$2^3F_4$ & 7.617 & --    	& 7.617 & --    		& --    \\
\hline
$3^3F_2$ & 7.946 & --    	& --    	& --    		& --\\
$3^3F_3$ & 7.954 & --    	& --    	& --    		& --\\
$3^1F_3$ & 7.953 &--     	& --    	& --    		& --\\
$3^3F_4$ & 7.956 & --    	& --    	& --    		& --
\\ \hline \end{tabular*}
\end{table}
\section{Decay properties}
The mass spectra of the hadronic states are experimentally determined through detection of energy and momenta of daughter particles in various decay channels. Generally, most phenomenological approaches obtain their model parameters like quark masses and confinement/Coulomb strength by fitting with the experimental ground states. So it becomes necessary for any phenomenological model to validate their fitted parameters through proper evaluation of various decay rates in general and annihilation rates in particular.
In the nonrelativistic limit, the decay properties are dependent on the wave function. In this section, we test our parameters and wave functions to determine various annihilation widths and electromagnetic transitions.
\subsection{Leptonic decay constants}
The leptonic decay constants of heavy quarkonia play very important role in understanding the weak decays. The matrix elements for leptonic decay constants of pseudoscalar and vector mesons are given by
\begin{eqnarray}
   \langle 0| \bar Q \gamma^\mu\gamma_5Q|P_\mu(k)\rangle  = i f_P k^\mu
\end{eqnarray}
\begin{eqnarray}
    \langle 0| \bar Q \gamma^\mu Q|P_\mu(k)\rangle  = i f_V M_V \epsilon^{*\mu}
\end{eqnarray}
where $k$ is the momentum of pseudoscalar meson, $\epsilon^{*\mu}$ is the polarization vector of meson. In the nonrelativistic limit, the decay constants of pseudoscalar and vector mesons are given by Van Royen-Weiskopf formula \cite{VanRoyen:1967}
\begin{eqnarray}\label{eq:decay_constant}
    f^2_{P/V} = \frac{3 |R_{nsP/V}(0)|^2}{\pi M_{nsP/V}} {\bar{C}}^2(\alpha_S).
\end{eqnarray}
Here the QCD correction factor ${\bar{C}}^2(\alpha_S)$ \cite{Braaten:1995,Berezhnoy:1996}
\begin{equation}\label{eq:decay_constant_correction}
\bar{C}^2(\alpha_S) = 1 - \frac{\alpha_s}{\pi} \left(\delta^{P,V} - \frac{m_Q - m_{\bar{Q}}}{m_Q + m_{\bar{Q}}} \text {ln} \frac{m_Q}{m_{\bar{Q}}}\right).
\end{equation}
With $\delta^P$ = 2 and $\delta^V$ = 8/3.
Using the above relations, we compute the leptonic decay constants $f_p$ and $f_v$ for charmonia, bottomonia and $B_c$ mesons. The results are listed in Tables \ref{tab:cc_fp} -- \ref{tab:bc_fv} in comparison with other models including LQCD.
\begin{table*}
    \caption{Pseudoscalar decay constant of charmonia (in MeV)}\label{tab:cc_fp}
   \begin{tabular*}{\textwidth}{@{\extracolsep{\fill}}cccccccc@{}}
   \hline
   State  &  $f_p$  & \cite{Patel:2008} & \cite{Krassnigg:2016} & \cite{Lakhina:2006} &   LQCD\cite{Becirevic:2013} & QCDSR \cite{Becirevic:2013} & PDG\cite{pdg2016}\\
   \hline
   $1S$  &  350.314 &  363 & 378 	& 402  	& 387(7)(2) 	& 309 $\pm$ 39	& 335 $\pm$ 75\\
   $2S$  &  278.447 &  275 & 82   	& 240	& --				&-- 					& -- \\
   $3S$  &  249.253 &  239 & 206 	& 193	& --				&-- 					& -- \\
   $4S$  &  231.211 &  217 & 87   	& --		& --				&-- 					& -- \\
   $5S$  &  218.241 &  202 & --    	& --		& --				&--\\
   $6S$  &  208.163 &  197 & --    	& --		& --				&-- 					& -- \\
 \hline \end{tabular*}
\end{table*}
\begin{table*}
 \caption{Vector decay constant of charmonia (in MeV)}\label{tab:cc_fv}
   \begin{tabular*}{\textwidth}{@{\extracolsep{\fill}}cccccccc@{}}
   \hline
   State  &  $f_v$  &   \cite{Patel:2008} & \cite{Krassnigg:2016}& \cite{Lakhina:2006}  &  LQCD\cite{Becirevic:2013} & QCDSR \cite{Becirevic:2013} & PDG\cite{pdg2016}\\
   \hline
   $1S$ &  325.876 &  338 & 411 & 393 &  418(8)(5) & 401 $\pm$ 46& 416 $\pm$ 6\\
   $2S$ &  257.340   &  254 & 155 & 293 &  --			& -- 					& 304 $\pm$ 4\\
   $3S$ &  229.857 &  220 & 188 & 258 &  --			& -- 					& -- \\
   $4S$ &  212.959 &  200 & 262 &	--	  &  --			& -- 					& -- \\
   $5S$ &  200.848 &  186 & --     &	--	  &  --			& -- 					& -- \\
   $6S$ &  191.459 &  175 & --     &  --	  & --				&-- 					& -- \\
 \hline \end{tabular*}
\end{table*}
\begin{table}
\caption{Pseudoscalar decay constant of bottomonia (in MeV)}\label{tab:bb_fp}
\begin{tabular*}{\columnwidth}{@{\extracolsep{\fill}}cccccc@{}}
\hline
State & $f_p$ &  \cite{Patel:2008} & \cite{Krassnigg:2016} & \cite{Pandya:2001} & \cite{Lakhina:2006} \\
\hline
$1S$ & 646.025 &  744 & 756 & 711& 599 \\
$2S$ & 518.803 &  577 & 285 & -- & 411 \\
$3S$ & 474.954 &  511 & 333 & -- & 354 \\
$4S$ & 449.654 &  471 & 40   & -- & -- \\
$5S$ & 432.072 &  443 & --    & -- & -- \\
$6S$ & 418.645 &  422 & --    & -- & --  \\
\hline \end{tabular*}
\end{table}
\begin{table*}
\caption{Vector decay constant of bottomonia (in MeV)}\label{tab:bb_fv}
\begin{tabular*}{\textwidth}{@{\extracolsep{\fill}}cccccccc@{}}
\hline
State & $f_v$ &  \cite{Patel:2008} & \cite{Krassnigg:2016} & \cite{Lakhina:2006} & \cite{Wang:2005} & LQCD\cite{Colquhoun:2014} & PDG\cite{pdg2016}\\
\hline
$1S$ & 647.250 &  706 & 707 &  665 & 498$\pm(20)$ & 649(31)	& 715 $\pm$ 5\\
$2S$ & 519.436 &  547 & 393 &  475 & 366$\pm(27)$ & 481(39)	& 498 $\pm$ 8\\
$3S$ & 475.440 &  484 & 9     &  418 & 304$\pm(27)$ & --     		& 430 $\pm$ 4\\
$4S$ & 450.066 &  446 & 20   &  388 & 259$\pm(22)$ &--    		& 336 $\pm$ 18\\
$5S$ & 432.437 &  419 & --     &  367 & 228$\pm(16)$ &--     		& --\\
$6S$ & 418.977 &  399 & --     &  351 & --            		 & --    		& --\\
 \hline \end{tabular*}
\end{table*}
\begin{table}
\caption{Pseudoscalar decay constant of $B_c$ meson (in MeV)}\label{tab:bc_fp}
\begin{tabular*}{\columnwidth}{@{\extracolsep{\fill}}ccccccc@{}}
\hline
State & $f_p$  &  \cite{Patel:2008} & \cite{Ebert:2002} & \cite{Gershtein:1995} &\cite{Eichten:1994} & \cite{Monteiro:2016} \\
\hline
$1S$ & 432.955 &  465 & 503&460$\pm(60)$ &500 & 554.125\\
$2S$ & 355.504 &  361 & --	& --					& --\\
$3S$ & 325.659 &  319 & --	& --					& --\\
$4S$ & 307.492 &  293 & --	& --					& --\\
$5S$ & 294.434 &  275 & --	& --					& --\\
$6S$ & 284.237 &  261 & --	& --					& --\\
\hline \end{tabular*}
\end{table}
\begin{table}
\caption{Vector decay constant of $B_c$ meson (in MeV)}\label{tab:bc_fv}
\begin{tabular*}{\columnwidth}{@{\extracolsep{\fill}}cccccc@{}}
\hline
State & $f_v$  & \cite{Patel:2008} & \cite{Ebert:2002} & \cite{Gershtein:1995} &\cite{Eichten:1994} \\
\hline
$1S$ & 434.642 &  435 & 433 & 460$\pm(60)$ 	& 500\\
$2S$ & 356.435 &  337 & --	& --						& --\\
$3S$ & 326.374 &  297 & --	& --						& --\\
$4S$ & 308.094 &  273 & --	& --						& --\\
$5S$ & 294.962 &  256 & --	& --						& --\\
$6S$ & 284.709 &  243 & --	& --						& --\\
 \hline \end{tabular*}
\end{table}
\subsection{Annihilation widths of heavy quarkonia}
Digamma, digluon and dilepton annihilation decay widths of heavy quarkonia are very important in understanding the dynamics of heavy quarks within the mesons. The measurement of digamma decay widths provides the information regarding the internal structure of meson.
The decay $\eta_c \to \gamma\gamma$, $\chi_{c0,2} \to \gamma\gamma$ was reported by CLEO-c \cite{Ecklund:2008}, \textit{BABAR} \cite{Lees:2010} and then BESIII \cite{Ablikim:2012} collaboration have reported high accuracy data. LQCD is found to underestimate the decay widths of $\eta_c \to \gamma\gamma$ and $\chi_{c0} \to \gamma\gamma$ when compared to experimental data \cite{Dudek:2006,ChenT:2016}.
Other approaches to attempt computation of annihilation rates of heavy quarkonia include NRQCD \cite{Bodwin:1994,Khan:1995,Schuler:1997,Bodwin:2006,Bodwin:2007},
relativistic quark model \cite{Ebert:2003gamma,Ebert:2003lepton},
effective Lagrangian \cite{Lansberg:2009,Lansberg:2006} and next-to-next-to leading order QCD correction to $\chi_{c0,2} \to \gamma \gamma$ in the framework of nonrelativistic QCD factorization \cite{Sang:2015}.

The meson decaying into digamma suggests that the spin can never be one \cite{Landau:1948,Yang:1950}. Corresponding digamma decay width of a pseudoscalar meson in nonrelativistic limit is given by Van Royen-Weiskopf formula \cite{VanRoyen:1967,Kwong:1988}
\begin{dmath}\label{eq:digamma}
    \Gamma_{n^1S_0 \to \gamma\gamma} = \frac{3 \alpha_e^2 e_Q^4 |R_{nsP}(0)|^2}{m_Q^2}  \left[1+ \frac{\alpha_s}{\pi} \left(\frac{\pi^2-20}{3}\right)\right]
\end{dmath}
\begin{dmath}\label{eq:digamma_p0}
\Gamma_{n^3P_0 \to \gamma\gamma} =\frac{27 \alpha_e^2 e_Q^4 |R_{nP}'(0)|^2}{ M_Q^4}\left[1+\frac{\alpha_s}{\pi}\left(\frac{3\pi^2-28}{9}\right)\right]
\end{dmath}
\begin{eqnarray}\label{eq:digamma_p2}
\Gamma_{n^3P_2 \to \gamma\gamma} = \frac{36 \alpha_e^2 e_Q^4 |R_{nP}'(0)|^2}{5 M_Q^4} \left[1 - \frac{16}{3} \frac{\alpha_s}{\pi}\right]
\end{eqnarray}
where the bracketed quantities are QCD next-to-leading order radiative corrections \cite{Kwong:1988,Barbieri:1981}.

Digluon annihilation of quarkonia is not directly observed in detectors as digluonic state decays into various hadronic states making it a bit complex to compute digluon annihilation widths from nonrelativistic approximations derived from first principles. The digluon decay width of pseudoscalar meson along with the QCD leading order radiative correction is given by \cite{Lansberg:2009,Kwong:1988,Barbieri:1981,Mangano:1995}
\begin{eqnarray}\label{eq:digluon}
    \Gamma_{n^1S_0 \to gg} = \frac{2 \alpha_s^2 |R_{nsP}(0)|^2}{3 m_Q^2} [1 + C_Q (\alpha_s/\pi)]
\end{eqnarray}
\begin{eqnarray}\label{eq:digluon_p0}
\Gamma_{n^3P_0 \to gg} = \frac{6 \alpha_s^2 |R_{nP}'(0)|^2}{m_Q^4} [1 + C_{0Q} (\alpha_s/\pi)]
\end{eqnarray}
\begin{dmath}\label{eq:digluon_p2}
\Gamma_{n^3P_2 \to gg} = \frac{4 \alpha_s^2|R_{nP}'(0)|^2}{5 m_Q^4} [1 + C_{2Q} (\alpha_s/\pi)]
\end{dmath}
Here, the coefficients in the bracket have values of $C_Q = 4.8$, $C_{0Q} = 9.5$, $C_{2Q} = -2.2$ for the charm quark and $C_Q = 4.4$, $C_{0Q} = 10.0$, $C_{2Q} = -0.1$ for the bottom quark \cite{Kwong:1988}.

The vector mesons have quantum numbers $1^{--}$ and can annihilate into dilepton. The dileptonic decay of vector meson along with one loop QCD radiative correction is given by \cite{VanRoyen:1967,Kwong:1988}
\begin{eqnarray}\label{eq:dilepton}
    \Gamma_{n^3S_1 \to \ell^+\ell^-} = \frac{4 \alpha_e^2 e_Q^2 |R_{nsV}(0)|^2}{M_{nsV}^2}  \left[1-\frac{16\alpha_s}{3\pi}\right]
\end{eqnarray}
Here, $\alpha_e$ is the electromagnetic coupling constant, $\alpha_s$ is the strong running coupling constant in Eq. (\ref{eq:running_coupling}) and $e_Q$ is the charge of heavy quark in terms of electron charge.
In above relations, $|R_{nsP/V} (0)|$ corresponds to the wave function of $S$-wave at origin for pseudoscalar and vector mesons while $|R_{nP}' (0)|$ is the derivative of $P$-wave wave function at origin.
The annihilation rates of heavy quarkonia are listed in Tables \ref{tab:cc_digamma} - \ref{tab:bb_di_lepton}.
\begin{table}
\caption{Digamma decay width of $S$ and $P$-wave charmonia (in keV)}\label{tab:cc_digamma}
\begin{tabular*}{\columnwidth}{@{\extracolsep{\fill}}ccccccc@{}}
\hline
State & $\Gamma_{\gamma\gamma}$   & \cite{Li:2009} & \cite{Ebert:2003gamma} & \cite{Lakhina:2006} & \cite{Kim:2004} &PDG \cite{pdg2016}\\
\hline
$1^1S_0$ &  7.231	& 8.5 	& 5.5   & 7.18	& 7.14$\pm$0.95 	&  5.1$\pm$0.4\\
$2^1S_0$ &  5.507 	& 2.4 	& 1.8   & 1.71	& 4.44$\pm$0.48 	& 2.15$\pm$1.58\\
$3^1S_0$ &  4.971 	& 0.88 	& --      & 1.21	& --						& -- \\
$4^1S_0$ &  4.688 	& -- 		& --      & --		& --						& --\\
$5^1S_0$ &  4.507 	& -- 		& --      & --		& --						& --\\
$6^1S_0$ &  4.377 	& -- 		& --      & --		& --						& --\\
$1^3P_0$ &  8.982	& 2.5 	& 2.9   & 3.28 	& --						& 2.34$\pm$0.19\\
$1^3P_2$ &  1.069 	& 0.31 	& 0.50 & --			& --						& 0.53$\pm$0.4\\
$2^3P_0$ &  9.111		& 1.7		& 1.9   & --			& --						& --\\
$2^3P_2$ &  1.084 	& 0.23 	& 0.52 & --			&  --						& --\\
$3^3P_0$ &  9.104 	& 1.2	  	& --      & --		& --						& -- \\
$3^3P_2$ &  1.0846 	& 0.17	& --      & --		& --						& -- \\
$4^3P_0$ &  9.076	& -- 		& --      & --		& --						& --\\
$4^3P_2$ &  1.080 	& -- 		& --      & --		& --						& --\\
$5^3P_0$ &  9.047	& -- 		& --      & --		& --						& -- \\
$5^3P_2$ &  1.077 	& -- 		& --      & --		& --						& --\\
\hline \end{tabular*}
\end{table}
\begin{table}
\caption{Digamma decay width of $S$ and $P$-wave bottomonia  (in keV)}\label{tab:bb_digamma}
\begin{tabular*}{\columnwidth}{@{\extracolsep{\fill}}ccccccc@{}}
\hline
State & $\Gamma_{\gamma\gamma}$ & \cite{Li:2009bb} & \cite{Godfrey:1985} & \cite{Ebert:2003gamma} & \cite{Lakhina:2006} & \cite{Kim:2004}\\
\hline
$1^1S_0$ & 0.387 	& 0.527 	& 0.214 	& 0.35 	& 0.23 	& 0.384 $\pm$ 0.047\\
$2^1S_0$ & 0.263 	& 0.263 	& 0.121 	& 0.15 	& 0.07 	& 0.191 $\pm$ 0.025\\
$3^1S_0$ & 0.229 	& 0.172 	& 0.906 	& 0.10 	& 0.04 	& --\\
$4^1S_0$ & 0.212 	& 0.105		& 0.755 	& --		& --		& --\\
$5^1S_0$ & 0.201 	& 0.121		& --			& --		& --		& --\\
$6^1S_0$ & 0.193 	& 0.050		& --			& --		& --		& --\\
$1^3P_0$ & 0.0196 	& 0.050 	& 0.0208 	& 0.038	& --		& --\\
$1^3P_2$ & 0.0052	& 0.0066	& 0.0051	& 0.008	& --		& --\\
$2^3P_0$ & 0.0195	& 0.037 	& 0.0227 	& 0.029	& --		& --\\
$2^3P_2$ & 0.0052 	& 0.0067	& 0.0062	& 0.006	& --		& --\\
$3^3P_0$ & 0.0194 	& 0.037		& --			& --		& --		& --\\
$3^3P_2$ & 0.0051 	& 0.0064	& --			& --		& --		& --\\
$4^3P_0$ & 0.0192 	& --			& --			& --		& --		& --\\
$4^3P_2$ & 0.0051 	& --			& --			& --		& --		& --\\
$5^3P_0$ & 0.0191 	& --			& --			& --		& --		& --\\
$5^3P_2$ & 0.0050	&--			& --			& --		& --		& --\\
\hline \end{tabular*}
\end{table}
\begin{table}
\caption{Digluon decay width of $S$ and $P$-wave charmonia  (in MeV)}\label{tab:cc_digluon}
\begin{tabular*}{\columnwidth}{@{\extracolsep{\fill}}ccccc@{}}
\hline
State & $\Gamma_{gg}$  & \cite{Patel:2015} & \cite{Kim:2004} & PDG \cite{pdg2016}\\
\hline
$1^1S_0$ &  35.909	& 22.37 & 19.60	& 26.7$\pm$3.0\\
$2^1S_0$ &  27.345 	& 16.74 & 12.1	& 14.7$\pm$0.7\\
$3^1S_0$ &  24.683 	& 14.30 & --		& --\\
$4^1S_0$ &  23.281 	& -- 		& --		& --\\
$5^1S_0$ &  22.379 	& --		& --		& --\\
$6^1S_0$ &  23.736 	& --		& --		& --\\
$1^3P_0$ &  37.919 	& 9.45 	& --		& 10.4$\pm$0.7\\
$1^3P_2$ &  3.974 	& 2.81 	& --		& 2.03$\pm$0.12\\
$2^3P_0$ &  38.462 	& 10.09	& --		& --\\
$2^3P_2$ &  4.034 	& 7.34 	& --		& --\\
$3^3P_0$ &  38.433 	& -- 		& --		& --\\
$3^3P_2$ &  4.028   	& -- 		& --		& --\\
$4^3P_0$ &  38.315 	& -- 		& --		& --\\
$4^3P_2$ &  4.016   	& -- 		& --		& --\\
$5^3P_0$ &  39.191 	& -- 		& --		& --\\
$5^3P_2$ &  4.003   	& -- 		& --		& --\\
\hline \end{tabular*}
\end{table}
\begin{table}
\caption{Digluon decay width of $S$ and $P$-wave bottomonia  (in MeV)}\label{tab:bb_digluon}
\begin{tabular*}{\columnwidth}{@{\extracolsep{\fill}}ccccc@{}}
\hline
State & $\Gamma_{gg}$ & \cite{Parmar:2010} & \cite{Kim:2004} & \cite{Gupta:1996gg} \\
\hline
$1^1S_0$ & 5.448 & 17.945 	& 6.98	& 12.46 \\
$2^1S_0$ & 3.710 & --			& 3.47	& --\\
$3^1S_0$ & 3.229 & --			& --		& --\\
$4^1S_0$ & 2.985 & --			& --		& --\\
$5^1S_0$ & 2.832 & --			& --		& -- \\
$6^1S_0$ & 2.274	& --			& --		& -- \\
$1^3P_0$ & 0.276	& 5.250 	& --		& 2.15\\
$1^3P_2$ & 0.073	& 0.822 	& --		& 0.22\\
$2^3P_0$ & 0.275	& --			& -- 		& --\\
$2^3P_2$ & 0.073	& --			&  -- 		& -- \\
$3^3P_0$ & 0.273	& --			&  -- 		& -- \\
$3^3P_2$ & 0.072	& --			&  -- 		& --\\
$4^3P_0$ & 0.271	& --			&  -- 		& --\\
$4^3P_2$ & 0.072	& --			&  -- 		& --\\
$5^3P_0$ & 0.269	& --			&  -- 		& --\\
$5^3P_2$ & 0.071 & --			&  -- 		& --\\
\hline \end{tabular*}
\end{table}
\begin{table}
    \caption{Dilepton decay width of charmonia (in keV)}\label{tab:cc_di_lepton}
   \begin{tabular*}{\columnwidth}{@{\extracolsep{\fill}}ccccccc@{}}
   \hline
   State  &  $\Gamma_{\ell^+\ell^-}$  & \cite{Shah:2012}  & \cite{Patel:2008} & \cite{Radford:2007} & \cite{Ebert:2003lepton} & PDG \cite{pdg2016}    \\
   \hline
   $1S$ &  2.925 & 4.95 & 6.99  & 1.89 	& 5.4 & 5.547 $\pm$ 0.14  \\
   $2S$ &  1.533 & 1.69 & 3.38  & 1.04 	& 2.4 & 2.359 $\pm$ 0.04   \\
   $3S$ &  1.091 & 0.96 & 2.31  & 0.77 	& --	 & 0.86 $\pm$ 0.07   \\
   $4S$ &  0.856 & 0.65 & 1.78  & 0.65		& --	 & 0.58 $\pm$ 0.07  \\
   $5S$ &  0.707 & 0.49 & 1.46  & --			& --	 & --\\
   $6S$ &  0.602 & 0.39 & 1.24  & --			& --	 & --
 \\ \hline \end{tabular*}
\end{table}
\begin{table}
\caption{Dilepton decay width of bottomonia (in  keV)}\label{tab:bb_di_lepton}
\begin{tabular*}{\columnwidth}{@{\extracolsep{\fill}}cccccccc@{}}
\hline
State & $\Gamma_{\ell^+\ell^-}$ & \cite{Shah:2012} & \cite{Radford:2009} & \cite{Patel:2008} & \cite{Ebert:2003lepton}  & \cite{Gonzalez:2003} & PDG\cite{pdg2016} \\
\hline
$1S$ & 1.098 & 1.20 & 1.33 	& 1.61 & 1.3 	 	& 0.98 	& 1.340 $\pm$ 0.018 \\
$2S$ & 0.670 & 0.52 & 0.62 	& 0.87 & 0.5 	 	& 0.41 	& 0.612 $\pm$ 0.011\\
$3S$ & 0.541 & 0.33 & 0.48 	& 0.66 & -- 	   	& 0.27 	& 0.443 $\pm$ 0.008\\
$4S$ & 0.470 & 0.24 & 0.40 	& 0.53 & -- 	   	& 0.20 	& 0.272 $\pm$ 0.029\\
$5S$ & 0.422 & 0.19 & --     	& 0.44 & -- 	 	&  0.16	& --\\
$6S$ & 0.387 & 0.16 & --     	& 0.39 & -- 	    & 0.12 	& --\\
 \hline \end{tabular*}
\end{table}
\subsection{Electromagnetic transition widths}
The electromagnetic transitions can be determined broadly in terms of electric and magnetic multipole expansions and their study can help in understanding the non-perturbative regime of QCD.
We consider the leading order terms i.e. electric ($E1$) and magnetic ($M1$) dipoles with selection rules $\Delta L = \pm 1$ and $\Delta S = 0$ for the $E1$ transitions while $\Delta L = 0$ and $\Delta S = \pm 1$ for $M1$ transitions.
We now employ the numerical wave function for computing the electromagnetic  transition widths among quarkonia and $B_c$ meson states in order to test parameters used in present work. For $M1$ transition, we restrict our calculations for transitions among $S$-waves only.
In the nonrelativistic limit, the radiative $E1$ and $M1$ widths are given by \cite{Eichten:1974,Eichten:1978,Brambilla:2010,Brambilla:2005,Li:2010}
\begin{dmath}
\label{eq:e1}
\Gamma(n^{2S+1}L_{iJ_i} \to {n'}^{2S+1}L_{fJ_f} + \gamma) = \frac{4 \alpha_e \langle e_Q\rangle ^2\omega^3}{3} (2 J_f + 1) S_{if}^{E1} |M_{if}^{E1}|^2
\end{dmath}
\begin{dmath}
\label{eq:m1}
\Gamma(n^3S_1 \to {n'}^{1}S_0+ \gamma) = \frac{\alpha_e \mu^2 \omega^3}{3} (2 J_f + 1) |M_{if}^{M1}|^2
\end{dmath}
where, mean charge content $\langle e_Q \rangle$ of the $Q\bar{Q}$ system, magnetic dipole moment $\mu$ and photon energy $\omega$ are given by
\begin{equation}
\langle e_Q \rangle = \left |\frac{m_{\bar{Q}} e_Q - e_{\bar{Q}} m_Q}{m_Q + m_{\bar{Q}}}\right |
\end{equation}
\begin{equation}
\mu = \frac{e_Q}{m_Q} - \frac{e_{\bar{Q}}}{m_{\bar{Q}}}
\end{equation}
and
\begin{equation}
\omega = \frac{M_i^2 - M_f^2}{2 M_i}
\end{equation}
respectively. Also the symmetric statistical factor is given by
\begin{equation}
S_{if}^{E1} = {\rm max}(L_i, L_f)
\left\{ \begin{array}{ccc} J_i & 1 & J_f \\ L_f & S & L_i \end{array}  \right\}^2\\.
\end{equation}
The matrix element $|M_{if}|$ for $E1$ and $M1$ transition can be written as
\begin{equation}
\left |M_{if}^{E1}\right | = \frac{3}{\omega} \left\langle f \left | \frac{\omega r}{2} j_0 \left(\frac{\omega r}{2}\right) - j_1 \left(\frac{\omega r}{2}\right) \right | i \right\rangle
\end{equation}
and
\begin{equation}
\left |M_{if}^{M1}\right | = \left\langle f\left | j_0 \left(\frac{\omega r}{2}\right)  \right | i \right\rangle
\end{equation}
The electromagnetic transition widths are listed in Tables \ref{tab:e1_cc} - \ref{tab:m1_bc} and also compared with experimental results as well as theoretical predictions.
\begin{table*}
\caption{$E1$ transition width of charmonia (in keV)}\label{tab:e1_cc}
\begin{tabular*}{\textwidth}{@{\extracolsep{\fill}}ccccccc@{}}
\hline
Transition & Present & \cite{Radford:2007} 	& \cite{Ebert:2002} & \cite{Li:2009} &\cite{Deng:2016cc}& PDG \cite{pdg2016}\\
\hline
$2^3S_1 \to 1^3P_0$ & 21.863 	& 45.0		& 51.7	& 74 		& 22 		& $29.8 \pm 1.5$\\
$2^3S_1 \to 1^3P_1$ & 43.292 	& 40.9		& 44.9	& 62 		& 42 		& $27.9 \pm 1.5$\\
$2^3S_1 \to 1^3P_2$ & 62.312 	& 26.5		& 30.9	& 43 		& 38 		& $26 \pm 1.5$\\
$2^1S_0 \to 1^1P_1$ & 36.197 	& 8.3			& 8.6 	& 146	& 49 		& --\\
\hline
$3^3S_1 \to 2^3P_0$ & 31.839 	& 87.3		& --		& -- 		& --		& --\\
$3^3S_1 \to 2^3P_1$ & 64.234 	& 65.7		& --		& -- 		& --		& --\\
$3^3S_1 \to 2^3P_2$ & 86.472 	& 31.6		& --		& -- 		& --		& --\\
$3^1S_0 \to 2^1P_1$ & 51.917 	& --			& --		& -- 		& --		& --\\
$3^3S_1 \to 1^3P_0$ & 46.872 	& 1.2			& --		& -- 		& --		& --\\
$3^3S_1 \to 1^3P_1$ & 107.088 & 2.5		& --		& -- 		& --		& --\\
$3^3S_1 \to 1^3P_2$ & 163.485 & 3.3		& --		& -- 		& --		& --\\
$3^1S_0 \to 1^1P_1$ & 178.312 & --			& --		& -- 		& --		& --\\
\hline
$1^3P_0 \to 1^3S_1$ & 112.030 & 142.2		& 161	& 167 	& 284 	& $119.5 \pm 8$\\
$1^3P_1 \to 1^3S_1$ & 146.317 & 287.0	& 333	& 354 	& 306 	& $295 \pm 13$\\
$1^3P_2 \to 1^3S_1$ & 157.225 & 390.6	& 448	& 473 	& 172 	& $384.2 \pm 16$\\
$1^1P_1 \to 1^1S_0$ & 247.971 & 610.0	& 723	& 764 	& 361 	& $357 \pm 280$\\
\hline
$2^3P_0 \to 2^3S_1$ & 70.400 	& 53.6		& --		& 61		& --		& --\\
$2^3P_1 \to 2^3S_1$ & 102.672 & 208.3	& --		& 103	& --		& --\\
$2^3P_2 \to 2^3S_1$ & 116.325 & 358.6		& --		& 225	& --		& --\\
$2^1P_1 \to 2^1S_0$ & 163.646 & --			& --		& 309	& --		& --\\
\hline
$2^3P_0 \to 1^3S_1$ & 173.324 & 20.8		& --		& 74		& --		& --\\
$2^3P_1 \to 1^3S_1$ & 210.958 & 28.4		& --		& 83		& --		& --\\
$2^3P_2 \to 1^3S_1$ & 227.915 & 33.2		& --		& 101	& --		& --\\
$2^1P_1 \to 1^1S_0$ & 329.384 & --			& --		& 134	& --		& --\\
\hline
$1^3D_1 \to 1^3P_0$ & 161.504	& --			& 423	& 486 	& 272 	& $172 \pm 30$\\
$1^3D_1 \to 1^3P_1$ & 93.775 	& --			& 142	& 150 	& 138 	& $70 \pm 17$\\
$1^3D_1 \to 1^3P_2$ & 5.722 	& --			& 5.8		& 5.8 	& 7.1  	& $\leq 21$\\
$1^3D_2 \to 1^3P_1$ & 165.176	& 317.3		& 297	& 342 	& 285 	& -- \\
$1^3D_2 \to 1^3P_2$ & 50.317 	& 65.7		& 62		& 70 		& 91 		& --\\
$1^3D_3 \to 1^3P_2$ & 175.212	& 62.7		& 252	& 284 	& 350 	& --\\
$1^1D_2 \to 1^1P_1$ & 205.93	& --			& 335	& 575	& 362 	& --\\
\hline \end{tabular*}
\end{table*}
\begin{table*}
\caption{$E1$ transition width of bottomonia (in keV)}\label{tab:e1_bb}
\begin{tabular*}{\textwidth}{@{\extracolsep{\fill}}ccccccc@{}}
\hline
Transition & Present & \cite{Radford:2007} & \cite{Ebert:2002} & \cite{Li:2009bb} & \cite{Deng:2016bb} & PDG \cite{pdg2016}  \\
\hline
$2^3S_1 \to 1^3P_0$ & 2.377	& 1.15	& 1.65 	& 1.67	& 1.09 & 1.22 $\pm$ 0.11\\
$2^3S_1 \to 1^3P_1$ & 5.689	& 1.87	& 2.57 	& 2.54	& 2.17 & 2.21 $\pm$ 0.19\\
$2^3S_1 \to 1^3P_2$ & 8.486	& 1.88	& 2.53 	& 2.62	& 2.62 & 2.29 $\pm$ 0.20\\
$2^1S_0 \to 1^1P_1$ & 10.181	& 4.17	& 3.25 	& 6.10	& 3.41 & -- \\
\hline
$3^3S_1 \to 2^3P_0$ & 3.330	& 1.67	& 1.65	& 1.83	& 1.21 & 1.20 $\pm$ 0.12\\
$3^3S_1 \to 2^3P_1$ & 7.936	& 2.74	& 2.65	& 2.96	& 2.61 & 2.56 $\pm$ 0.26\\
$3^3S_1 \to 2^3P_2$ & 11.447	& 2.80	& 2.89	& 3.23	& 3.16 & 2.66 $\pm$ 0.27\\
$3^3S_1 \to 1^3P_0$ & 0.594	& 0.03	& 0.124	& 0.07	& 0.097 & 0.055 $\pm$ 0.010\\
$3^3S_1 \to 1^3P_1$ & 1.518	& 0.09	& 0.307	& 0.17	& 0.0005 & 0.018 $\pm$ 0.010\\
$3^3S_1 \to 1^3P_2$ & 2.354	& 0.13	& 0.445	& 0.15	& 0.14 & 0.20 $\pm$ 0.03\\
$3^1S_0 \to 1^1P_1$ & 3.385	& 0.03	& 0.770	& 1.24	& 0.67 & --\\
$3^1S_0 \to 2^1P_1$ & 13.981	& --		& 3.07	& 11.0	& 4.25 & --\\
\hline
$1^3P_2 \to 1^3S_1$ & 57.530	& 31.2	& 29.5	& 38.2	& 31.8 & --\\
$1^3P_1 \to 1^3S_1$ & 54.927	& 27.3 	& 37.1	& 33.6	& 31.9 & --\\
$1^3P_0 \to 1^3S_1$ & 49.530	& 22.1	& 42.7	& 26.6	& 27.5 & --\\
$1^1P_1 \to 1^1S_0$ & 72.094	& 37.9	& 54.4	& 55.8	& 35.8 & --\\
\hline
$2^3P_2 \to 2^3S_1$ & 28.848	& 16.8	& 18.8	& 18.8	& 15.5 & 15.1 $\pm$ 5.6\\
$2^3P_1 \to 2^3S_1$ & 26.672	& 13.7	& 15.9	& 15.9	& 15.3 & 19.4 $\pm$ 5.0\\
$2^3P_0 \to 2^3S_1$ & 23.162	& 9.90	& 11.7	& 11.7	& 14.4 & --\\
$2^1P_1 \to 2^1S_0$ & 35.578	& --		&  23.6	& 24.7	& 16.2 & --\\
\hline
$2^3P_2 \to 1^3S_1$ & 29.635	& 7.74	& 8.41	& 13.0	& 12.5 & 9.8 $\pm$ 2.3\\
$2^3P_1 \to 1^3S_1$ & 28.552	& 7.31	& 8.01	& 12.4	& 10.8 & 8.9 $\pm$ 2.2\\
$2^3P_0 \to 1^3S_1$ & 26.769	& 6.69	& 7.36	& 11.4	& 5.4 & --\\
$2^1P_1 \to 1^1S_0$ & 34.815	& --		& 9.9		& 15.9	& 16.1 & --\\
\hline
$1^3D_1 \to 1^3P_0$ & 9.670	& --		& 24.2	& 23.6	& 19.8 & --\\
$1^3D_1 \to 1^3P_1$ & 6.313	& --		& 12.9	&12.3	& 13.3 & --\\
$1^3D_1 \to 1^3P_2$ & 0.394	& --		& 0.67	& 0.65	& 1.02 & --\\
$1^3D_2 \to 1^3P_1$ & 11.489	& 19.3	& 24.8	& 23.8	& 21.8 & --\\
$1^3D_2 \to 1^3P_2$ & 3.583	& 5.07	& 6.45	& 6.29	& 7.23 & --\\
$1^3D_3 \to 1^3P_2$ & 14.013	& 21.7	& 26.7	& 26.4	& 32.1 & --\\
$1^1D_2 \to 1^1P_1$ & 14.821	& --		& 30.2	& 42.3	& 30.3 & --
\\ \hline \end{tabular*}
\end{table*}
\begin{table}
\caption{$E1$ transition width of $B_c$ meson (in keV)}\label{tab:e1_bc}
\begin{tabular*}{\columnwidth}{@{\extracolsep{\fill}}ccccc@{}}
\hline
Transition & Present & \cite{Ebert:2002} & \cite{Godfrey:2004} & \cite{Devlani:2014}\\
\hline
$2^3S_1 \to 1^3P_0$ & 4.782	& 5.53	& 2.9		& 0.94	\\
$2^3S_1 \to 1^3P_1$ & 11.156	& 7.65	& 4.7		& 1.45	\\
$2^3S_1 \to 1^3P_2$ & 16.823	& 7.59	& 5.7		& 2.28	\\
$2^1S_0 \to 1^1P_1$ & 18.663	& 4.40	& 6.1		& 3.03	\\
\hline
$3^3S_1 \to 2^3P_0$ & 7.406	& --		& --		& --		\\
$3^3S_1 \to 2^3P_1$ & 17.049	& --		& --		& --		 \\
$3^3S_1 \to 2^3P_2$ & 25.112	& --		& --		& --		 \\
$3^3S_1 \to 1^3P_0$ & 6.910	& --		& --		& --		 \\
$3^3S_1 \to 1^3P_1$ & 17.563	& --		& --		& --		 \\
$3^3S_1 \to 1^3P_2$ & 27.487	& --		& --		& --		 \\
$3^1S_0 \to 1^1P_1$ & 38.755	& --		& --		& --		 \\
$3^1S_0 \to 2^1P_1$ & 27.988	& --		& --		& --		 \\
\hline
$1^3P_2 \to 1^3S_1$ & 55.761	& 122	& 83		& 64.24	\\
$1^3P_1 \to 1^3S_1$ & 53.294	& 87.1	& 11		& 51.14	\\
$1^3P_0 \to 1^3S_1$ & 46.862	& 75.5	& 55		& 58.55	\\
$1^1P_1 \to 1^1S_0$ & 71.923	& 18.4	& 80		& 72.28	\\
\hline
$2^3P_2 \to 2^3S_1$ & 41.259	& 75.3	& 55		& 64.92	 \\
$2^3P_1 \to 2^3S_1$ & 38.533	& 45.3	& 45		& 50.40	 \\
$2^3P_0 \to 2^3S_1$ & 38.308	& 34.0	& 42		& 55.05	 \\
$2^1P_1 \to 2^1S_0$ & 52.205	& 13.8	& 52		& 56.28	 \\
\hline
$2^3P_2 \to 1^3S_1$ & 60.195	& --		& 14		& --		 \\
$2^3P_1 \to 1^3S_1$ & 57.839	& --		& 5.4		& --		 \\
$2^3P_0 \to 1^3S_1$ & 52.508	& --		& 1.0		& --		 \\
$2^1P_1 \to 1^1S_0$ & 74.211	& --		& 19		& --		 \\
\hline
$1^3D_1 \to 1^3P_0$ & 44.783	& 133	& 55		& --		\\
$1^3D_1 \to 1^3P_1$ & 28.731	& 65.3	& 28		& --		\\
$1^3D_1 \to 1^3P_2$ & 1.786	& 3.82	& 1.8		& --		\\
$1^3D_2 \to 1^3P_1$ & 51.272	& 139	& 64		& --		 \\
$1^3D_2 \to 1^3P_2$ & 16.073	& 23.6	& 15		& --		 \\
$1^3D_3 \to 1^3P_2$ & 60.336 	& 149	& 78		& --		 \\
$1^1D_2 \to 1^1P_1$ & 66.020	& 143	& 63		& --		
\\ \hline \end{tabular*}
\end{table}

\begin{table}
\caption{$M1$ transition width of charmonia (in keV)}\label{tab:m1_cc}
\begin{tabular*}{\columnwidth}{@{\extracolsep{\fill}}ccccccc@{}}
\hline
Transition & Present & \cite{Radford:2007} & \cite{Ebert:2002}& \cite{Deng:2016cc} & \cite{Bhaghyesh:2011} & PDG \cite{pdg2016}\\
\hline
$1^3S_1 \to 1^1S_0$ & 2.722 & 2.7	& 1.05	& 2.39 	& 3.28 	& 1.58 $\pm$ 0.37\\
$2^3S_1 \to 2^1S_0$ & 1.172 & 1.2	& 0.99	& 0.19 	& 1.45 	& 0.21 $\pm$ 0.15\\
$2^3S_1 \to 1^1S_0$ & 7.506 & 0.0	& 0.95	& 7.80 	& --		& 1.24 $\pm$ 0.29\\
$3^3S_1 \to 3^1S_0$ & 9.927 & -- 		& --		& 0.088	& --		&  --
\\ \hline \end{tabular*}
\end{table}
\begin{table}
\caption{$M1$ transition width of bottomonia (in eV)}\label{tab:m1_bb}
\begin{tabular*}{\columnwidth}{@{\extracolsep{\fill}}ccccccc@{}}
\hline
Transition & Present & \cite{Radford:2007} & \cite{Ebert:2002} & \cite{Deng:2016bb} & \cite{Bhaghyesh:2011}& PDG \cite{pdg2016}\\
\hline
$1^3S_1 \to 1^1S_0$ & 37.668	& 4.0		& 5.8 	& 10		& 15.36 & --\\
$2^3S_1 \to 2^1S_0$ & 5.619	& 0.05	& 1.40	& 0.59	& 1.82	& --\\
$2^3S_1 \to 1^1S_0$ & 77.173	& 0.0 	& 6.4		& 66		& --		& 12.5 $\pm$ 4.9\\
$3^3S_1 \to 3^1S_0$ & 2.849	& --		& 0.8		& 3.9		& --		& --\\
$3^3S_1 \to 2^1S_0$ & 	36.177	& --		& 1.5		& 11		& --		& $\leq$ 14\\
$3^3S_1 \to 1^1S_0$ & 76.990	& --		& 10.5	& 71		& --		& 10 $\pm$ 2\\
 \hline \end{tabular*}
\end{table}
\begin{table}
\caption{$M1$ transition width of $B_c$ meson (in eV)}\label{tab:m1_bc}
\begin{tabular*}{\columnwidth}{@{\extracolsep{\fill}}ccccc@{}}
\hline
Transition & Present & \cite{Ebert:2002} & \cite{Godfrey:2004} &\cite{Devlani:2014} \\
\hline
$1^3S_1 \to 1^1S_0$ & 53.109 		& 33		& 80 		& 2.2		\\
$2^3S_1 \to 2^1S_0$ & 21.119 		& 17		& 10 		& 0.014	\\
$2^3S_1 \to 1^1S_0$ & 481.572 	& 428	& 600 	& 495	\\
$2^1S_0 \to 1^3S_1$ & 568.346		& 488	& 300 	& 1092	
\\ \hline \end{tabular*}
\end{table}

\subsection{Weak decays of $B_c$ mesons}
\label{sec:weak_decays}
The decay modes of $B_c$ mesons are different from charmonia and bottomonia because of the inclusion of different flavor quarks. Their decay properties are very important probes for the weak interaction as $B_c$ meson decays only through weak decays, therefore have relatively quite long life time. The pseudoscalar state can not decay via strong or electromagnetic decays because  of this flavor asymmetry.

In the spectator model \cite{El-Hady:1999}, the total decay width of $B_c$ meson can be broadly classified into three classes. (i) Decay of $b$ quark considering $c$ quark as a spectator, (ii) Decay of $c$ quark considering $b$ quark as a spectator and (iii) Annihilation channel $B_c \rightarrow \ell^+ \nu_\ell$. The total width is given by
\begin{equation}\label{eq:weak_Bc}
\Gamma (B_c \rightarrow X) = \Gamma (b \rightarrow X) + \Gamma(c \rightarrow X) + \Gamma (Anni)
\end{equation}
In the calculations of total width we have not considered the interference among them as all these decays lead to different channel. In the spectator approximation, the inclusive decay width of $b$ and $c$ quark is given by
\begin{equation}
\Gamma (b \rightarrow X) = \frac{9 G_F^2 |V_{cb}|^2 m_b^5}{192 \pi^3}
\end{equation}
\begin{equation}
\Gamma (c \rightarrow X) = \frac{9 G_F^2 |V_{cs}|^2 m_c^5}{192 \pi^3}
\end{equation}
\begin{equation}
\Gamma (Anni) = \frac{G_F^2}{8 \pi} |V_{bc}|^2 f_{B_c}^2 M_{B_c} m_q^2 \left(1-\frac{m_q^2}{M_{B_c^2}}\right)^2 C_q
\end{equation}
Where $C_q = 3 |V_{cs}|$ for $D_s$ mesons and $m_q$ is the mass of heaviest fermions. $V_{cs}$ and  $V_{cb}$ are the CKM matrices and we have taken the value of CKM matrices from the PDG. $G_f$ is the Fermi coupling constant. Here we have used the model quark masses, $B_c$ meson mass and decay constants for the computation of total width. Here we compute the decay width  of $B_c$ meson using Eq. (\ref{eq:weak_Bc}) and corresponding life time. The computed life time comes out to be $0.539 \times 10^{-12}$ s which is in very good agreement with the world averaged mean life time $(0.507 \pm 0.009) \times 10^{-12}$ s \cite{pdg2016}.
\section{Numerical results and discussion}
\label{sec:results}
Having determined the confinement strengths and quark masses, we are now in position to present our numerical results. We first compute the mass spectra of heavy quarkonia and $B_c$ meson.
In most of the potential model computations, the confinement strength is fixed by experimental ground state masses for $c\bar{c}$, $b\bar{b}$ and $c\bar{b}$ independently. We observe here that the confinement strength $A$ for $B_c$ meson is arithmetic mean of those for $c\bar c$ and $b\bar b$ which discards the need to introduce additional confinement strength parameter for computation of $B_c$ spectra. Similar approach has been used earlier within QCD potential model \cite{Gupta:1996}. Using model parameters and numerical wave function we compute the various decay properties of heavy quarkonia and $B_c$ mesons namely leptonic decay constants, annihilation widths and electromagnetic transitions.
In Tables \ref{tab:cc_sp_mass} and \ref{tab:cc_df_mass}, we present our result for charmonium mass spectra. We compare our results with PDG data \cite{pdg2016}, lattice QCD \cite{Kalinowski:2015} data, relativistic quark model \cite{Ebert:2011}, nonrelativistic quark model \cite{Deng:2016cc,Lakhina:2006}, QCD relativistic functional approach \cite{Fischer:2014}, relativistic potential model \cite{Radford:2007} and nonrelativistic potential models \cite{Shah:2012,Patel:2015,Li:2009,Barnes:2005}.
Our results for $S$-wave are in excellent agreement with the experimental data \cite{pdg2016}.
We determine the mass difference for $S$-wave charmonia i.e. $M_{{{\mathit J / \psi}}} - M_{\eta_c}$ = 105 $MeV$ and $M_{\psi(2S)} - M_{\eta_c (2S)}$ = 79 $MeV$ while that from experimental data are 113 $MeV$ and 47 $MeV$ respectively \cite{pdg2016}.
Our results for $P$-waves are also consistent with the PDG data \cite{pdg2016} as well as LQCD \cite{Kalinowski:2015} with less than 2\% deviation. Since experimental/LQCD results are not available for $P$-wave charmonia beyond $n=2$ states, we compare our results with the relativistic quark model \cite{Ebert:2011} and it is also observed to have 1-2 \% deviation throughout the spectra.
For charmonia, only $1 P$ states are available and for $2 P$ only one state is available namely $\chi_{c2}$. Our results for $1 P$ and $2 P$ states are also satisfactory.
We also list the mass spectra of $D$ and $F$ wave and find it to be consistent with the theoretical predictions. Overall, computed charmonium spectra is consistent with PDG and other theoretical models.

In Tables \ref{tab:bb_sp_mass} and \ref{tab:bb_df_mass}, we compare our results of bottomonium spectra with PDG data \cite{pdg2016}, relativistic quark model \cite{Ebert:2011,Godfrey:2015}, nonrelativistic quark model \cite{Deng:2016bb},QCD relativistic functional approach \cite{Fischer:2014}, relativistic potential model \cite{Radford:2009}, nonrelativistic potential models \cite{Shah:2012,Li:2009bb} and covariant constituent quark model \cite{Segovia:2016}.
Similarly for $S$-wave bottomonia, up to $n = 3$ vector states are known experimentally and for pseudoscalar states, only $n = 1$ and $2$ are available.
Our results for $\Upsilon (1S)$ and $\Upsilon (3S)$ are in good agreement with the PDG data while for $\Upsilon (2S)$, $\Upsilon (4S)$ and $\Upsilon (5S)$, slight deviation (within 1\%) is observed.
Our results for $\eta_b(1S)$ and $\eta_b(3S)$ also match well with less than 0.5 \% deviation.
We obtain $M_{\Upsilon(1S)} - M_{\eta_b}$ = 35 $MeV$ and for $M_{\Upsilon(2S)} - M_{\eta_b (2S)}$ = 24 $MeV$ against the PDG data of 62 $MeV$ and 24 $MeV$ respectively.
For $P$-wave, 1$P$ and 2$P$ states are reported and for 3$P$, only $\chi_{b1}$ is reported. Our results for 1$P$ bottomonia deviate by $\simeq$ 0.3\% from the experimental results but for 2$P$, they are quite satisfactory and deviating by 0.2 \% only from the PDG data. Our result for $\Upsilon(1D)$ also agrees well with the experimental data with 0.8 \% deviation. The $F$-wave mass spectra is also in good agreement with the theoretical predictions.
Looking at the comparison with PDG data Ref. \cite{pdg2016} and relativistic quark model Ref. \cite{Ebert:2011}, present quarkonium mass spectra deviate less than 2 \% for charmonia  and less than 1 \% for bottomonia.

We now employ the quark masses and confinement strengths used for computing mass spectra of quarkonia to predict the spectroscopy of $B_c$ mesons without introducing any additional parameter. Our results are tabulated in Tables \ref{tab:bc_sp_mass} and \ref{tab:bc_df_mass}. For $B_c$ mesons, only $0^{-+}$ states are experimentally observed for $n=1$ and $2$ and our results are in very good agreement with the experimental results with less than 0.3 \% error.

We note here that the masses of orbitally excited states (especially $n=1$ states) of charmonia are systematically lower than the other models and experimental data. This tendency decreases as one moves to higher $n$ states. Absence of similar trend in case of $B_c$ and bottomonia systems suggests that relativistic treatment might improve the results in lower energy regime of charmonia.

\begin{figure*}[htbp]
\includegraphics[width=0.32\textwidth]{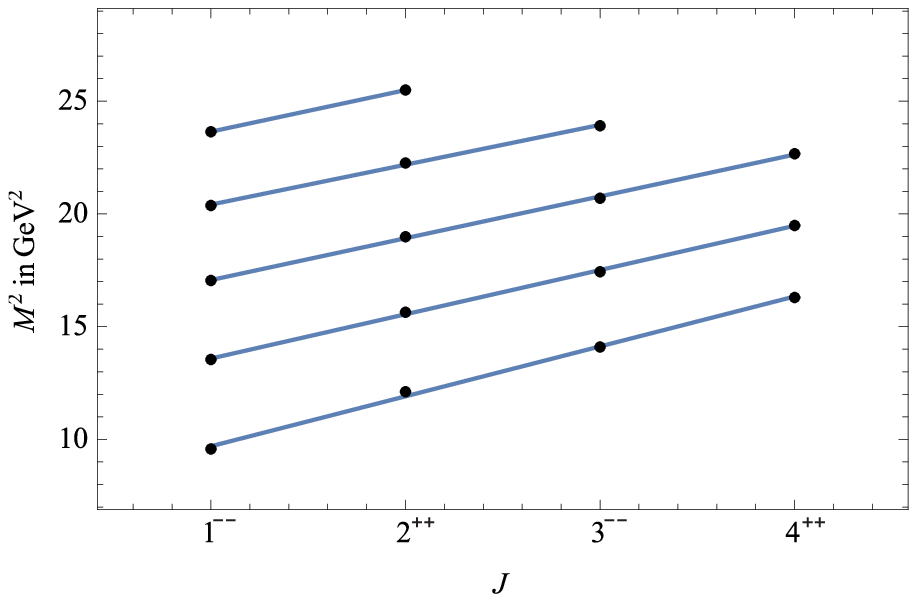}
\hfill\includegraphics[width=0.32\textwidth]{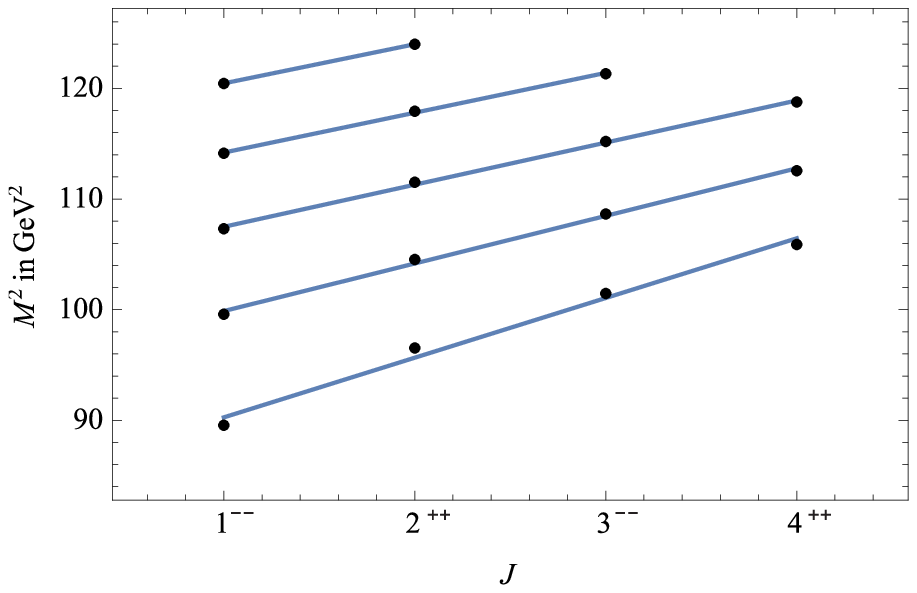}
\hfill\includegraphics[width=0.32\textwidth]{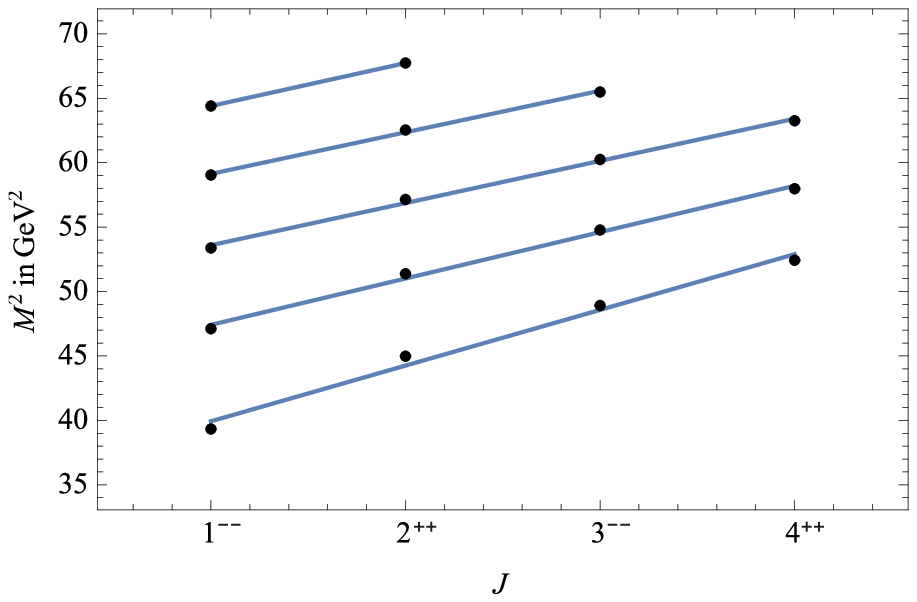}
\caption{Parent and daughter Regge trajectories ($J,M^2$) for charmonia (left), bottomonia (middle) and $B_c$ (right) mesons with natural parity ($P = (-1)^J$).}
\label{fig:natural}
\end{figure*}

\begin{figure*}[htbp]
\includegraphics[width=0.32\textwidth]{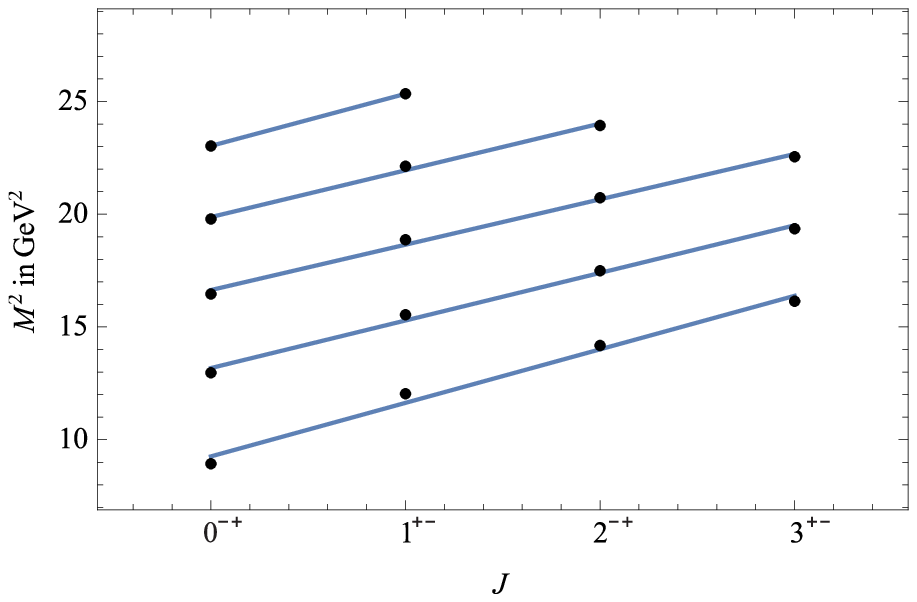}
\hfill\includegraphics[width=0.32\textwidth]{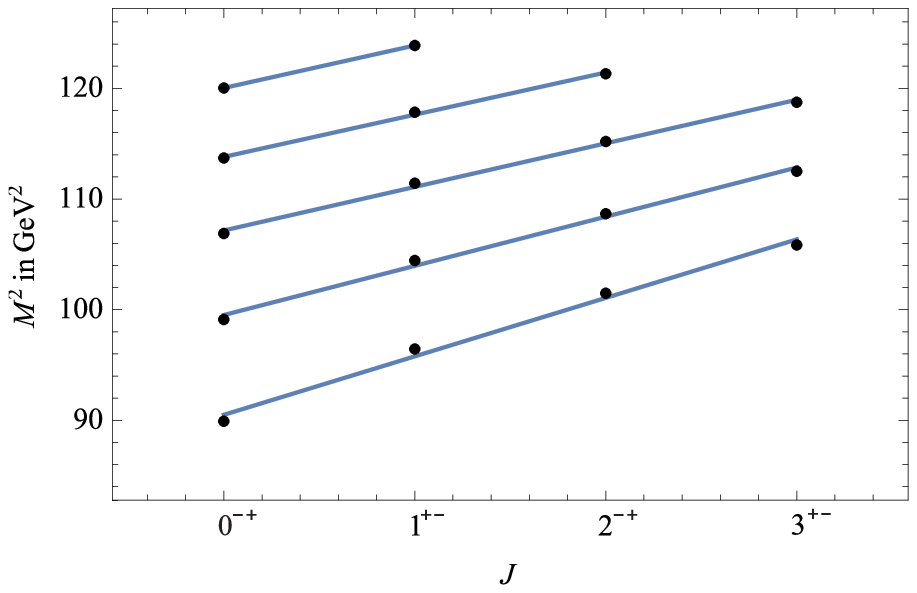}
\hfill\includegraphics[width=0.32\textwidth]{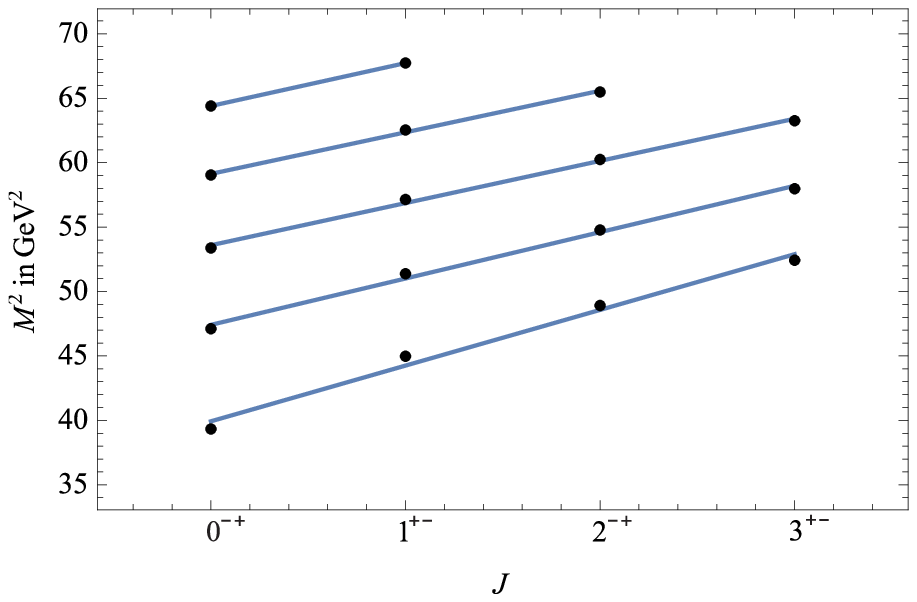}
\caption{Parent and daughter Regge trajectories ($J,M^2$) for charmonia (left), bottomonia (middle) and $B_c$ (right) mesons with unnatural parity ($P = (-1)^{J+1}$).}
\label{fig:unnatural}
\end{figure*}

\begin{figure*}[htbp]
\includegraphics[width=0.32\textwidth]{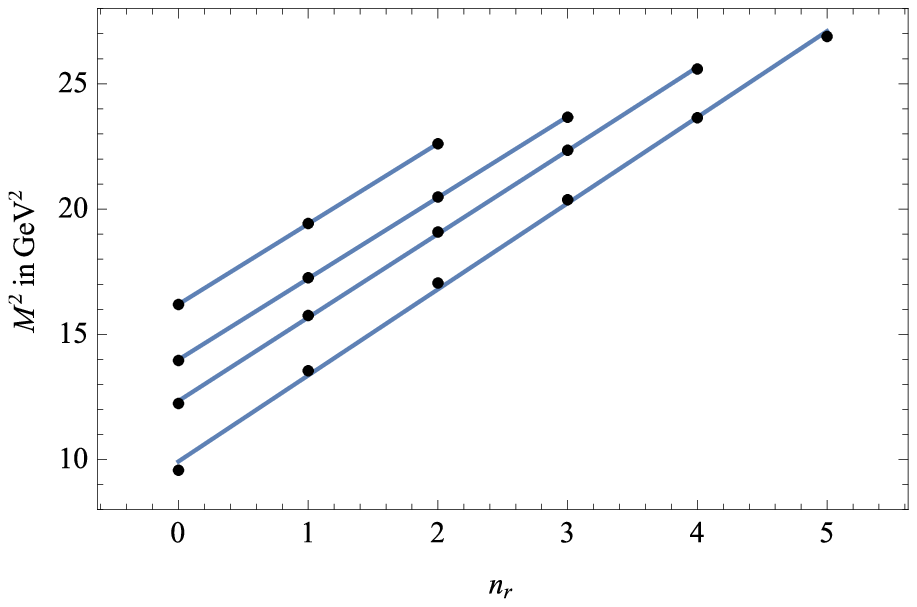}
\hfill\includegraphics[width=0.32\textwidth]{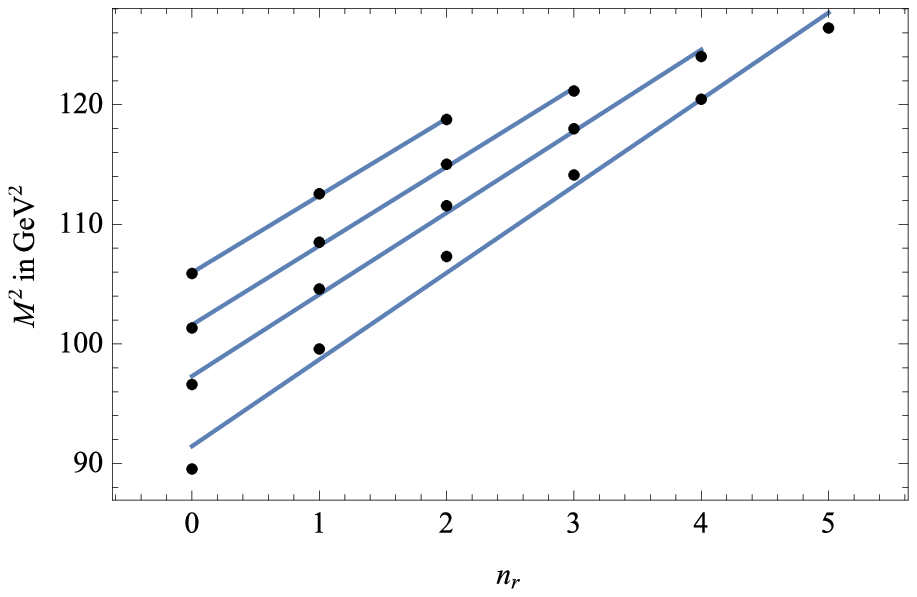}
\hfill\includegraphics[width=0.32\textwidth]{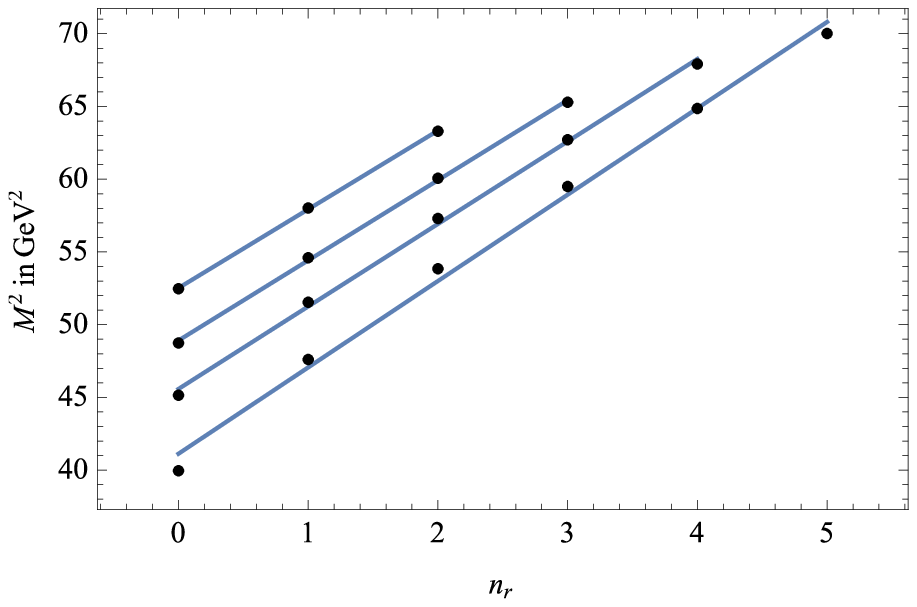}
\caption{Parent and daughter Regge trajectories ($n_r \to M^2$) for charmonia (left), bottomonia (middle) and $B_c$ (right) mesons}
\label{fig:nr}
\end{figure*}

Using the mass spectra of heavy quarkonia and $B_c$ meson, we plot the Regge trajectories in $(J, M^2)$ and $(n_r, M^2)$ planes where $n_r$ = $n - 1$. We use the following relations \cite{Ebert:2011}
\begin{eqnarray}
J = \alpha M^2 + \alpha_0\\
n_r = \beta M^2 + \beta_0
\end{eqnarray}
where $\alpha$, $\beta$ are slopes and $\alpha_0$, $\beta_0$ are the intercepts that can be computed using the methods given in Ref. \cite{Ebert:2011}. In Figs. \ref{fig:natural}, \ref{fig:unnatural} and \ref{fig:nr}, we plot the Regge trajectories.
Regge trajectories from present approach and relativistic quark model \cite{Ebert:2011} show similar trend i.e. for charmonium spectra, the computed mass square fits very well to a linear trajectory and found to be almost parallel and equidistant in both the planes.
Also, for bottomonia and $B_c$ mesons, we observe the nonlinearity in the parent trajectories. The nonlinearity increases as we go from $c\bar{b}$ to $b\bar{b}$ mesons indicating increasing contribution from the inter-quark interaction over confinement.
According to the first principles of QCD, while the one-gluon-exchange interaction gives rise to employment of Coulomb potential with a strength proportional to the strong running coupling constant at very short distances, nonperturbative effect like confinement becomes prominent at larger distances. Charmonium belongs to neither purely nonrelativistic nor the relativistic regime where chiral symmetry breaking is more significant from physics point of view.  Though Lattice QCD calculations in the quenched approximation have suggested a linearly increasing potential in the confinement range \cite{Dudek:2007,Meinel:2009,Burch:2009,Liu:2012,McNeile:2012,Daldrop:2011,Kawanai:2013,Kawanai:2011,Burnier:2015,Kalinowski:2015,Burnier:2016}, a specific form of interaction  potential in the full range is not yet known. At short distances relativistic effects are more important as they give rise to quark-antiquark pairs from the vacuum that in turn affect the nonrelativistic Coulomb interaction in the presence of sea quarks. The mass spectra of quarkonia is not sensitive to these relativistic effects at short distances. However, the decay properties show significant difference with inclusion of relativistic corrections. We have used the most accepted available correction terms for computation of decay properties \cite{Lansberg:2009,Kwong:1988,Barbieri:1981,Mangano:1995} that improves the results significantly in most cases.

Using the potential parameters and numerical wave function, we compute the various decay properties of heavy quarkonia. We first compute the leptonic decay constants of pseudoscalar and vector mesons and our numerical results are tabulated in Tables \ref{tab:cc_fp} -- \ref{tab:bc_fv}. For the case of charmonia, our results are higher than those using LQCD and QCDSR \cite{Becirevic:2013}. In order to overcome this discrepancy, we include the QCD correction factors given in Ref. \cite{Braaten:1995} and the results are tabulated in Table \ref{tab:cc_fp} and Table \ref{tab:cc_fv}. After introducing the correction factors our results match with PDG, LQCD and QCDSR \cite{Becirevic:2013} along with other theoretical models. We also compute the decay constants for excited $S$- wave charmonia and we found that our results are consistent with the other theoretical predictions.
We also compute the decay constants of bottomonia and $B_c$ mesons. In this case, our results match with other theoretical predictions without incorporating the relativistic corrections. In the case of vector decay constants of bottomonia, our results are very close to experimental results as well as those obtained in LQCD Ref. \cite{Colquhoun:2014}. For the decay constants of $B_c$ mesons, we compare our results with nonrelativistic potential models \cite{Patel:2008,Monteiro:2016}.

Next we compute the digamma, digluon and dilepton decay widths using the relations Eqs. (\ref{eq:digamma})--(\ref{eq:digluon}). Where the bracketed quantities are the first order radiative corrections to the decay widths. We compare our results with the available experimental results. We also compare our results with the theoretical models such as screened potential model \cite{Li:2009,Li:2009bb}, Martin-like potential model \cite{Shah:2012}, relativistic quark model (RQM) \cite{Ebert:2003gamma,Ebert:2003lepton}, heavy quark spin symmetry \cite{Lansberg:2006}, relativistic Salpeter model \cite{Kim:2004} and other theoretical data.

Tables \ref{tab:cc_digamma} and \ref{tab:bb_digamma} we present our results for digamma decay widths for charmonia and bottomonia.
Our results for $\Gamma(\eta_c \to \gamma\gamma)$ and $\Gamma(\eta_c(2S) \to \gamma\gamma)$ are higher than the experimental results. Experimental observation of the two photon decays of pseudoscalar states are considered as an important probe for identification of flavour as well as internal structure of mesons. The first order radiative correction (bracketed terms in Eq. (\ref{eq:digamma})) was utilized to incorporate the difference
and it is observed that our results along with the correction match with the experimental results \cite{pdg2016}. We also compute the digamma decay width of excited charmonia.
Our results for $P$-wave charmonia are higher than that of screened potential model \cite{Li:2009} and relativistic quark model \cite{Ebert:2003gamma}.
Our results for $\Gamma(\eta_b \to \gamma\gamma)$ match quite well with the experimental data while computed $\Gamma(\eta_b(2S) \to \gamma\gamma)$ value is overestimated when compared with the PDG data.
For the excited state of $S$-wave bottomonia, our results fall in between those obtained in screened potential model \cite{Li:2009bb} and relativistic quark model with linear confinement \cite{Godfrey:2015}. The scenario is similar with $P$-wave bottomonia and charmonia.

Di-gluon decay has substantial contribution to hadronic decay of quarkonia below $c\bar c$ and $b\bar b$ threshold. In Tables \ref{tab:cc_digluon} and \ref{tab:bb_digluon} we represent our results for digluon decay width of charmonia and bottomonia respectively. Our results for $\Gamma(\eta_c \to gg)$ match perfectly with the PDG data \cite{pdg2016} but in the case of $\Gamma(\eta_c (2S) \to gg)$ our result is higher than the PDG data. We also compare the results obtained with that of the relativistic Salpeter method \cite{Kim:2004} and an approximate potential model \cite{Patel:2015}. It is seen from Table \ref{tab:cc_digluon} that the relativistic corrections provide better results in case of $P$-wave charmonia where as that for bottomonia are underestimated in present calculations when compared to relativistic QCD potential model \cite{Gupta:1996gg} and power potential model \cite{Parmar:2010}. As the experimental data of digluon annihilation of bottomonia are not available, the validity of either of the approaches can be validated only after observations in forthcoming experiments.

We present the result of dilepton decay widths in the Table \ref{tab:cc_di_lepton} and \ref{tab:bb_di_lepton} and it is observed that our results matches with the PDG data \cite{pdg2016} upto $n$ = 3 for both charmonia and bottomonia. The contribution of the correction factor is more significant in the excited states with compared to that in the ground states of the quarkonia, indicating different dynamics in the intermediate quark-antiquark distance. Our results are also in good accordance with the other theoretical models.

We present our results of $E1$ transitions in Tables \ref{tab:e1_cc} - \ref{tab:e1_bc} in comparison with theoretical attempts such as relativistic potential model \cite{Radford:2007}, quark model \cite{Ebert:2002}, nonrelativistic screened potential model \cite{Deng:2016bb,Li:2009,Li:2009bb}. We also compare our results of charmonia transitions with available experimental results.
Our result for $\Gamma (\psi(2S) \to \chi_{cJ}(1P) + \gamma)$ is in good agreement with the experimental result for $J = 0$ but our results for $J=1,2$ are higher than the PDG data.
Our results also agree well for the transition $\Gamma (\chi_{c2}(1P) \to J/\psi + \gamma)$. We also satisfy the experimental constraints for the transition $\Gamma(1^3D_1 \to \chi_{cJ} + \gamma)$ for $J=0,1,2$. Our results share the same range with the results computed in other theoretical models.
The $E1$ transitions of bottomonia agree fairly well except for the channel $\Gamma(\Upsilon (3S) \to \chi_{bJ}(3P))$, where our results are higher than the experimental results.
The comparison of our results of $E1$ transitions in $B_c$ mesons with relativistic quark model \cite{Ebert:2002,Godfrey:2004} and power potential model \cite{Devlani:2014} are found to be in good agreement.
In Tables \ref{tab:m1_cc} - \ref{tab:m1_bc}, we present our results of $M1$ transitions and also compared with relativistic potential model \cite{Radford:2007}, quark model \cite{Ebert:2002,Godfrey:2015}, nonrelativistic screened potential model \cite{Deng:2016cc,Deng:2016bb}, power potential \cite{Devlani:2014} as well as with available experimental results.
Our results of $\Gamma({\text n}\psi \to {\text n'}\eta_c + \gamma)$ are in very good agreement with the PDG data as well with the other theoretical predictions. Computed $M1$ transitions in $B_c$ mesons are also within the results obtained from theoretical predictions.
The computed $M1$ transition of bottomonia are found to be higher than the PDG data and also theoretical predictions.

\section{Conclusion}
In this article, we have reported a comprehensive study of heavy quarkonia in the framework of nonrelativistic potential model considering linear confinement with least number of free model parameters such as confinement strength and quark mass. They are fine tuned to obtain the corresponding spin averaged ground state masses of charmonia and bottomonia determined from experimental data. The parameters are then used to predict the masses of excited states. In order to compute mass spectra of orbitally excited states, we incorporate contributions from the spin dependent part of confined one gluon exchange potential perturbatively.

Our results are found to be consistent with available PDG data, LQCD, relativistic quark model and other theoretical potential models. We also compute the digamma, digluon and dilepton decay widths of heavy quarkonia using nonrelativistic Van-Royen Weiskopf formula. The first order radiative corrections in calculation of these decays provide satisfactory results for the charmonia while no such correction is needed in case of bottomonia for being purely nonrelativistic system.
We employ our parameters in computation of $B_c$ spectroscopy employing the quark masses and mean value of confinement strength of charmonia and bottomonia and our results are also consistent with the PDG data. We also compute the weak decays of $B_c$ mesons and the computed life time is also consistent with the PDG data. It is interesting to note here that despite having a $c$ quark, the nonrelativistic calculation of $B_c$ spectroscopy is in  very good agreement with experimental and other theoretical models.
\section*{Acknowledgment}
J.N.P. acknowledges the support from the University Grants Commission of India under Major Research Project F.No.42-775/2013(SR).
\clearpage
\bibliography{apssamp}
\end{document}